\documentclass[%
reprint,
superscriptaddress,
 amsmath,amssymb,
 aps,
 prc,
]{revtex4-2}

\usepackage{graphicx}
\usepackage{dcolumn}
\usepackage{bm}
\usepackage{hyperref}

\begin{document}

\newcommand{\alum}{$^{25}\mathrm{Al}$}
\newcommand{\phos}{$^{26}\mathrm{P}$}
\newcommand{\sili}{$^{25}\mathrm{Si}$}
\newcommand{\magn}{$^{24}\mathrm{Mg}$}

\preprint{APS/123-QED}
\title{\textbf{Precision $\beta$-delayed charged-particle emission spectroscopy at FRIB:\\Proof of principle with the $\beta$-decay of $^{25}$Si}}

\author{E. A. M. Jensen}
 \email[Contact author: ]{erik.jensen@chalmers.se}
 \affiliation{
 Institut for Fysik og Astronomi, Aarhus Universitet, Aarhus, 8000, Denmark
 }%
 \affiliation{
 Institutionen för fysik och astronomi, Chalmers tekniska högskola, 412 96, Göteborg, Sweden
 }

\author{J. M. Eder}
 \affiliation{
 Institut for Fysik og Astronomi, Aarhus Universitet, Aarhus, 8000, Denmark
 }%
 \affiliation{
 GSI Helmholtzzentrum für Schwerionenforschung GmbH, 64291, Darmstadt, Germany
 }

\author{P. H. Pedersen}
 \affiliation{
 Institut for Fysik og Astronomi, Aarhus Universitet, Aarhus, 8000, Denmark
 }

\author{A. Adams}
 \affiliation{
 Department of Physics and Astronomy, Michigan State University, East Lansing, Michigan 48824, USA
 }%
 \affiliation{
 Facility for Rare Isotope Beams, Michigan State University, East Lansing, Michigan 48824, USA
 }

\author{M. J. G. Borge}
 \affiliation{
 Instituto de Estructura de la Materia, CSIC, Madrid, 28006, Spain
 }

\author{B. A. Brown}
 \affiliation{
 Department of Physics and Astronomy, Michigan State University, East Lansing, Michigan 48824, USA
 }%
 \affiliation{
 Facility for Rare Isotope Beams, Michigan State University, East Lansing, Michigan 48824, USA
 }

\author{J. Dopfer}
 \affiliation{
 Department of Physics and Astronomy, Michigan State University, East Lansing, Michigan 48824, USA
 }%
 \affiliation{
 Facility for Rare Isotope Beams, Michigan State University, East Lansing, Michigan 48824, USA
 }

\author{H. O. U. Fynbo}
 \affiliation{
 Institut for Fysik og Astronomi, Aarhus Universitet, Aarhus, 8000, Denmark
 }

\author{B. S. O. Johansson}
 \affiliation{
 Institut for Fysik og Astronomi, Aarhus Universitet, Aarhus, 8000, Denmark
 }

\author{B. Jonson}
 \affiliation{
 Institutionen för fysik och astronomi, Chalmers tekniska högskola, 412 96, Göteborg, Sweden
 }

\author{M. Madurga}
 \affiliation{
 Department of Physics and Astronomy, University of Tennessee, Knoxville, Tennessee 37996, USA
 }

\author{J. S. Nielsen}
 \affiliation{
 Institut for Fysik og Astronomi, Aarhus Universitet, Aarhus, 8000, Denmark
 }

\author{K. Riisager}
 \affiliation{
 Institut for Fysik og Astronomi, Aarhus Universitet, Aarhus, 8000, Denmark
 }

\author{C. S. Sumithrarachchi}
 \affiliation{
 Facility for Rare Isotope Beams, Michigan State University, East Lansing, Michigan 48824, USA
 }

\author{L. J. Sun}
 \affiliation{
 Facility for Rare Isotope Beams, Michigan State University, East Lansing, Michigan 48824, USA
 }

\author{O. Tengblad}
 \affiliation{
 Instituto de Estructura de la Materia, CSIC, Madrid, 28006, Spain
 }

\author{L. E. Weghorn}
 \affiliation{
 Department of Physics and Astronomy, Michigan State University, East Lansing, Michigan 48824, USA
 }%
 \affiliation{
 Facility for Rare Isotope Beams, Michigan State University, East Lansing, Michigan 48824, USA
 }

\author{T. Wheeler}
 \affiliation{
 Department of Physics and Astronomy, Michigan State University, East Lansing, Michigan 48824, USA
 }%
 \affiliation{
 Facility for Rare Isotope Beams, Michigan State University, East Lansing, Michigan 48824, USA
 }%

\author{C. Wrede}
 \affiliation{
 Department of Physics and Astronomy, Michigan State University, East Lansing, Michigan 48824, USA
 }%
 \affiliation{
 Facility for Rare Isotope Beams, Michigan State University, East Lansing, Michigan 48824, USA
 }

\date{\today}

\begin{abstract}%
We report on the $\beta$-delayed proton and $\gamma$-ray emission from \sili{}, measured at the Facility for Rare Isotope Beams (FRIB).
Low-energy \sili{} ions extracted from the Advanced Cryogenic Gas Stopper were implanted into a thin carbon foil surrounded by a compact, highly segmented array of silicon detector telescopes and two high-purity germanium detectors.
This setup provides high-resolution charged-particle spectroscopy, establishing a proof of principle for precision stopped-beam decay studies at FRIB.
We reconstruct the \sili{} decay scheme, resolving new high-energy proton transitions and determining the feeding to excited states in \magn{}.
The observation of spectral interference patterns enables firm spin and parity assignments for highly excited states in \alum{}.
The \sili{} $\beta$-strength distribution is extracted and compared with large-scale shell-model calculations.
\end{abstract}

\maketitle

\section{Introduction}
The study of $\beta$-delayed charged-particle emission in proton-rich nuclei provides a sensitive probe of nuclear structure, offering direct access to Gamow-Teller strength distributions, isospin symmetry breaking, and the properties of highly excited unbound states \cite{Bor13}.
However, precision measurements of these decays are frequently limited by experimental challenges, including beam impurities, $\beta$-particle energy summing, and degraded resolution due to thick stopping targets.
Overcoming these limitations requires the delivery of purified, low-energy radioactive ion beams coupled with highly granular detection systems.

The combination of the Advanced Rare Isotope Separator (ARIS) \cite{Por23, Fuk23} and the Advanced Cryogenic Gas Stopper (ACGS) \cite{Lun20, Rin21} at the Facility for Rare Isotope Beams (FRIB) provides these exact capabilities.
During the first $\beta$-decay campaign in the FRIB Gas Stopping Area, a compact array of silicon detector telescopes and high-purity germanium detectors were deployed around a thin carbon catcher foil.
The primary objective of this campaign is the precise $\beta$-delayed 2-proton spectroscopy of the proton-halo candidates $^{22}$Al \cite{Jen26} and \phos{} \cite{Per16}, the former of which being the prime candidate, due to its lowest Coulomb barrier, for direct 2-proton emission \cite{Bro90}.

The beams developed by ARIS for studying $^{22}$Al and \phos{} were beams consisting of isotonic chains with $N=9$ and $N=11$, respectively, $N$ being the neutron number.
In the $N=11$ isotonic chain, both \phos{} and \sili{} were present.
It was known in advance that \sili{} would become a primary beam contaminant following thermalization in the ACGS, due to the chemical properties of silicon and phosphorus which caused \sili{} to emerge from the ACGS along with \phos{} as a molecular ligand of $^{25}\mathrm{SiH}$.
Because it was not possible to mass-separate the two species between the ACGS and the experimental setup, four hours of beam time were allocated to measuring the decay of \sili{} alone (employing a molecular beam of $^{25}\mathrm{SiO}_2$) to accurately characterize the background for the \phos{} experiment.

While the main features of the decay of \sili{} have been established in previous experiments \cite{Hat92, Rob93, Tho04, Sun21, Ste24}, the rich and well-documented nuclear structure of the $\beta$-decay daughter, \alum{}, and the $\beta$-delayed proton daughter, \magn{}, makes $\beta$-delayed protons from \sili{} an ideal in-beam calibration standard.
Consequently, the \sili{} data provide an excellent proof of principle for the capabilities of the present detection setup.
The combination of high yields from the ACGS and the compact, background-suppressed geometry of the detection array allows for charged-particle spectroscopy of unprecedented resolution in this mass region.

In this paper, we exploit these capabilities to provide an updated and highly precise description of the decay of \sili{}.
We identify new high-energy proton transitions and quantify the previously ambiguous feeding to excited states in $^{24}$Mg.
Furthermore, we utilize proton-line interference patterns to firmly assign spins and parities to highly excited states in $^{25}$Al.
Significant differences in observed proton emission intensities from deformed initial states in \alum{} to different deformed final states in \magn{} are also used to assign spins to highly excited states in \alum{} based on the results from \cite{Rii24}.
Finally, the revised decay scheme is used to extract the total \sili{} $\beta$-strength distribution.
We are able to extract the Gamow-Teller strength up to near the electron capture Q-value limit.
The results will be compared to state-of-the-art shell-model calculations in order to understand $\beta$-quenching in proton-rich nuclei.

Overall, this work demonstrates that the stopped beam infrastructure at FRIB, combining in-flight beams from ARIS with the purification capabilities of the ACGS, enables charged-particle spectroscopy at a resolution and sensitivity that reveals previously unresolved proton branches and nuclear structure.

In the following section describing the experimental details, we demonstrate the proof of principle for precision $\beta$-delayed charged-particle emission spectroscopy in the Gas Stopping Area at FRIB.
Details necessary to employ \sili{} as an in-beam calibration source in future experiments are also provided.

\section{Experimental details}\label{sec:experiment}

\begin{figure}
\includegraphics[width=0.98\columnwidth]{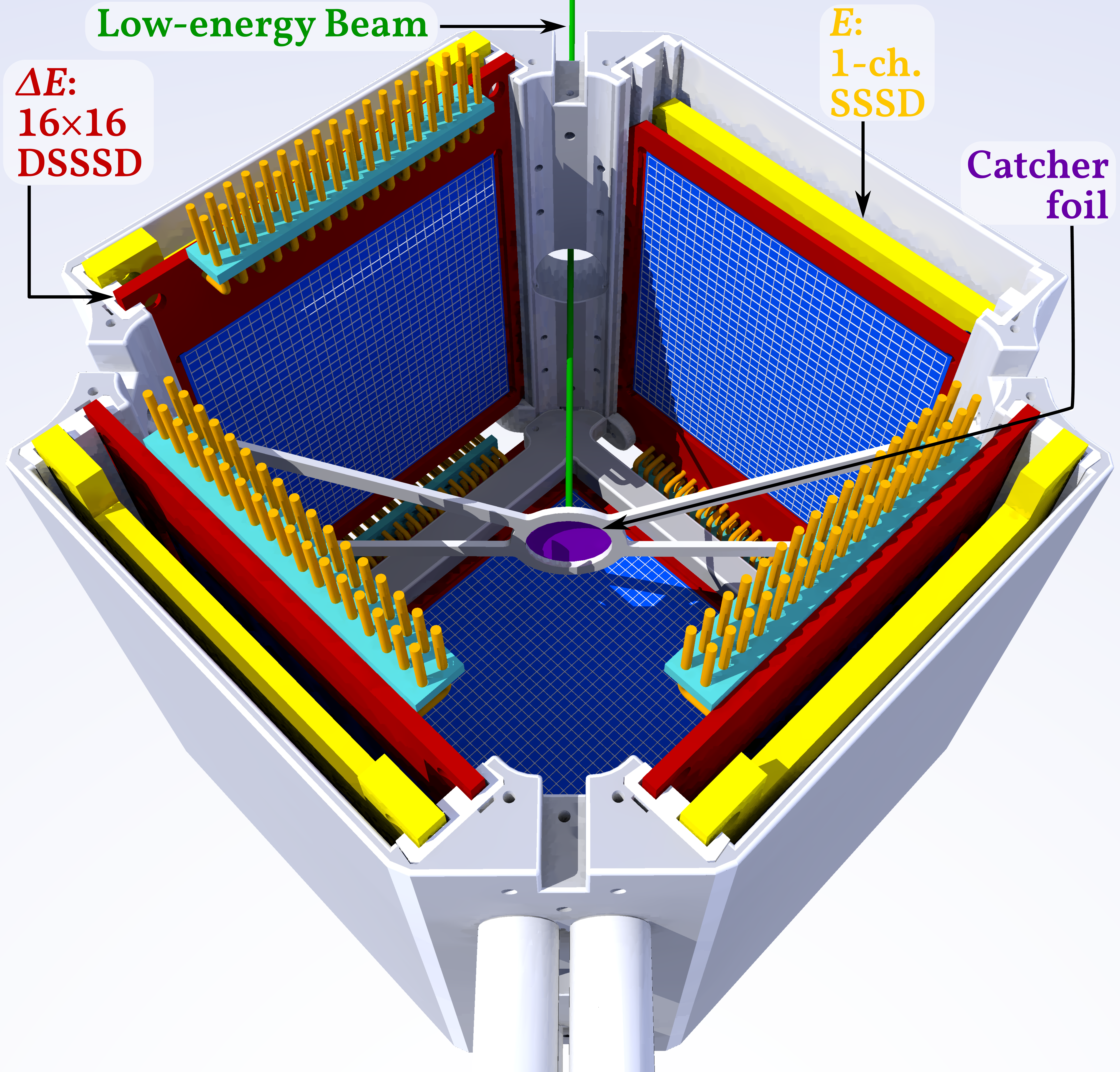}
\caption{\label{fig:cube}%
Render of silicon detector holder with detectors and catcher foil mounted.
A low-energy beam enters the holder from the top of the figure through a hole and is stopped in the catcher foil.
The detectors are placed in $\Delta E$-$E$ telescope configurations, with the $\Delta E$ detectors being 16$\times$16 Double-sided Silicon Strip Detectors (DSSDs) and the $E$ detectors being 1-channel Single-sided Silicon Detectors (SSDs).
The top lid, housing a 16$\times$16 DSSD, has been removed in this render.
}
\end{figure}

\begin{figure}
\includegraphics[width=0.98\columnwidth]{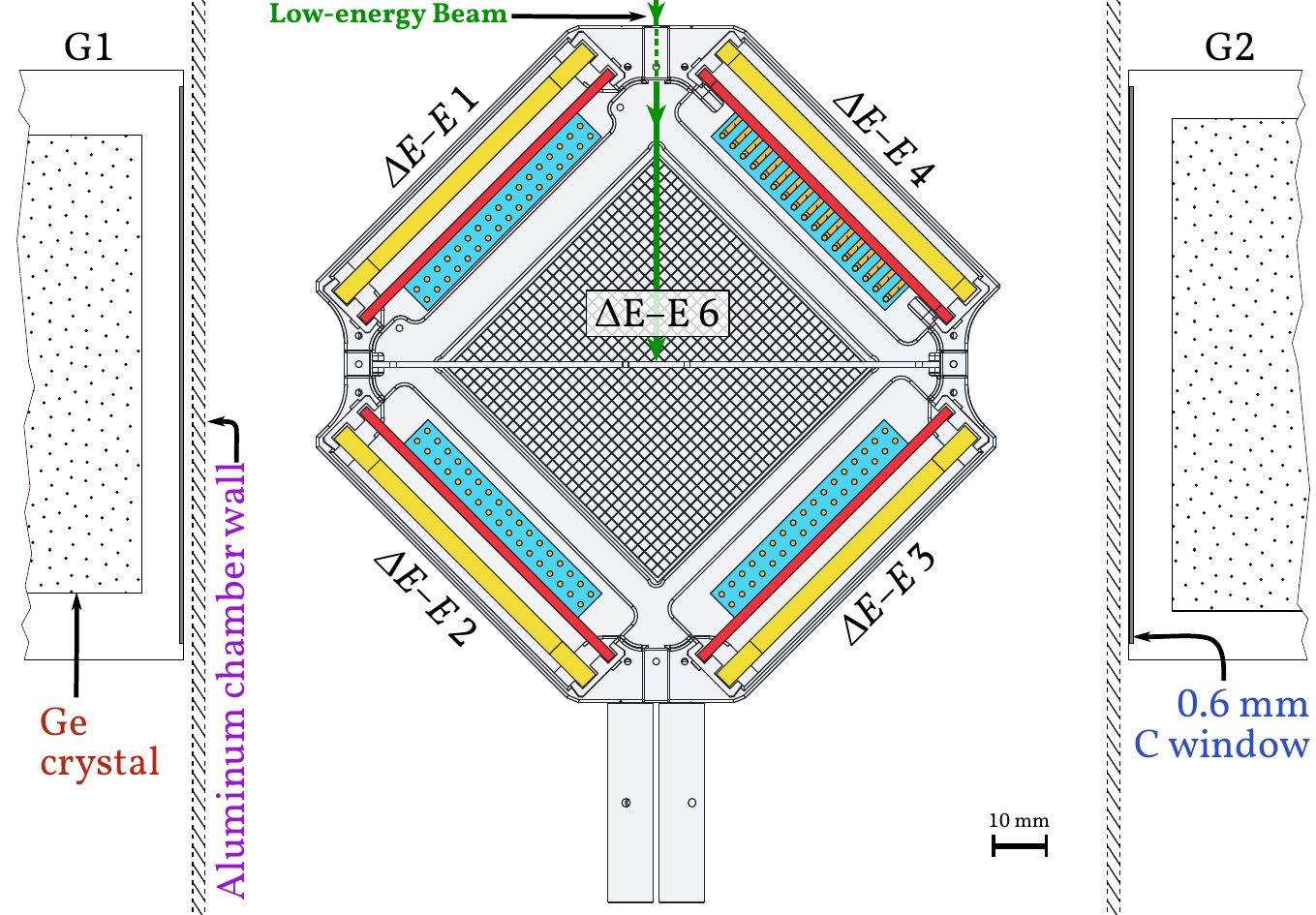}
\caption{\label{fig:chamber}%
Top-down schematic view of the silicon detector holder of Fig. \ref{fig:cube} placed inside an aluminum vacuum chamber and flanked by two high-purity germanium detectors, G1 and G2, outside the chamber.
The numbering of the $\Delta E$ and $E$ detectors is the same as that listed in Table \ref{tab:silicons}.
As in Fig. \ref{fig:cube}, the top lid has been removed.
The top lid houses '$\Delta E$-5', but has no backing $E$ detector.
}
\end{figure}

A 5 kW primary beam of $^{36}$Ar was accelerated to 210 MeV/u before impinging on a 8.07 mm carbon production target.
The resulting in-flight cocktail beam was momentum-to-charge-separated in ARIS, reducing the beam energy to 106 MeV/u.
Subsequently, the beam was guided to the Gas Stopping Area for further momentum compression, thermalization and purification.
A pure 30 keV beam of $^{25}\mathrm{SiO}_{2}$ was extracted from the ACGS and implanted into a thin carbon foil surrounded by a compact array of silicon detector telescopes; see Fig. \ref{fig:cube}.
Just outside of the vacuum chamber, containing the silicon detector telescopes, were two high-purity germanium detectors from the LIBRA setup \cite{Sun25}, flanking the chamber; see Fig. \ref{fig:chamber}.

Throughout the four hours of measurement time, the number of $\beta$-delayed protons from \sili{} detected in the silicon detectors was $7.8\times10^{5}$.
Taking into account the $\beta$-delayed proton branching ratio of \sili{}, $39.3(15)$\% \cite{Sun21}, and the energy-dependent solid angle coverage of the silicon detectors, the implantation rate at the detection setup is estimated to be around 350--400 particles per second.

The following subsection describes the employed detection setup in more detail.
Further details on the machine conditions at FRIB during the experiment can be found in \cite{Jens24}.

\subsection{Detection setup}\label{subsec:detectors}

The 30 keV beam from the ACGS was delivered to the detection setup at the General Purpose Line in the Stopped Beam Area at FRIB.
The vacuum chamber of the detection setup was an aluminum tube with an inner diameter of 155 mm and an outer diameter of 159 mm.

In the middle of the vacuum chamber, a silicon detector holder was placed.
The render in Fig. \ref{fig:cube} shows the silicon detector holder.
It has six faces in a shape resembling a cube and is made of 3D-printed aluminum.
The edge of the cube facing upstream has a hole through which the beam entered the cube.
In the middle of the cube, a holder for a thin catcher foil was placed, and on each of the six faces of the cube, silicon detector telescopes could be placed, facing the catcher foil.

The catcher foil was a 32 µg/cm$^{2}$, i.e. $150$ nm, natural carbon foil from Arizona Carbon Foil.
A 30 keV beam of $^{25}\mathrm{SiO}_{2}$ was stopped in the catcher foil, effectively yielding a beam energy of $25/57\times30\,\mathrm{keV} = 13\,\mathrm{keV}$ as the ions dissociated on impact with the catcher foil.
A \sili{} beam of this energy has a projected range of 17 nm in natural carbon \cite{Zie10}.
As a result, the particles emitted from the foil and detected upstream will suffer a smaller energy loss in escaping the catcher foil as compared to the particles detected downstream.
Range straggling will spread out the beam, but not beyond 50 nm.
Hence, no beam particles are expected to penetrate the entire catcher foil.

Flanking the vacuum chamber were two Canberra Extended Range (XtRa) Coaxial Germanium Detectors of model GX10020, from the LIBRA setup \cite{Sun25}.
These detectors face the thin catcher foil in the center of the silicon detector holder, just outside the vacuum chamber, as is shown in Fig. \ref{fig:chamber}.
The data acquisition (DAQ) system which was used for the experiment was unfortunately not well-suited to deliver the necessary signal shaping parameters to achieve the high energy resolutions and high detection efficiencies which the germanium detectors were otherwise capable of.
Nevertheless, the germanium detectors have been used to great benefit in discerning the decay scheme of \sili{}, which will be demonstrated below.

For the silicon detector telescopes used in the experiment, the $\Delta E$ detectors are of W1-type from Micron Semiconductor with window type 9G, as described in \cite{Ten04,Vin21}.
The detectors are double-sided silicon strip detectors (DSSDs) with $16 \times 16$ strips spanning an area of $50 \times 50\,\textrm{mm}^2$.
The $E$ detectors of the silicon detector telescopes are single-channel single-sided silicon detectors (SSDs) of MSX25-type from Micron Semiconductor, adapted to fit on the same chip as the W1-type.
The $E$ detectors have the same area dimensions as the $\Delta E$ detectors, $50 \times 50\,\textrm{mm}^2$.

\begin{table*}
\caption{\label{tab:silicons}%
Thicknesses, energy thresholds, punch through energies (from SRIM \cite{Zie10}) and solid angle coverages of silicon detectors.
The numbering of the $\Delta E$ and $E$ detectors is the same as in Fig. \ref{fig:chamber}.
Solid angle coverages are calculated with respect to the beam implantation spot.
Note that the $\Delta E$ detectors deliberately have a slightly smaller solid angle coverage than their backing $E$ detectors due to the exclusion of their outermost strips; see text.
The estimated systematic uncertainty of the solid angle coverages is 0.2\%.
}
\begin{ruledtabular}
\begin{tabular}{r|ccccccccccc}
                                    & $\Delta E$-1 & $\Delta E$-2 & $\Delta E$-3 & $\Delta E$-4 & $\Delta E$-5 & $\Delta E$-6 & $E$-1   & $E$-2   & $E$-3   & $E$-4    & $E$-6  \\
\colrule
Active layer thickness (µm)         & 53$^b$       & 69$^b$       & 59$^b$       & 300$^c$      & 1002$^c$     & 67$^b$       & 500$^c$ & 500$^c$ & 505$^c$ & 1498$^c$ & 500$^c$\\
Punch through energy$^a$ (MeV)      & 2.1--2.5     & 2.5--2.9     & 2.3--2.6     & 6.0--6.9     &              & 2.4--2.8     &         &         &         &          &        \\
Low-energy threshold (keV)          & 200--300     & 150--250     & 150--200     & 100          & 150          & 100          & 280     & 340     & 190     & 400      & 250    \\
Solid angle coverage (\% of $4\pi$) & 6.77         & 6.20         & 6.70         & 6.90         & 5.24         & 5.24         & 7.22    & 7.12    & 7.10    & 7.66     & 5.69
\end{tabular}
\end{ruledtabular}
\begin{flushleft}
$^a$Energy required for proton to punch through detector at angles of incidence with respect to normal $\theta=0$\textdegree~and 35\textdegree.\\
$^b$Active layer thickness estimated based on methods described in appendix \ref{app:thickness-estimation}.\\
$^c$Active layer thickness from detector's specification sheet.\\

\end{flushleft}
\end{table*}

In the horizontal plane of the silicon detector holder are 4 $\Delta E$-$E$ telescopes.
The bottom face of the detector holder also contains a W1-MSX25 ($\Delta E$-$E$) telescope, and the top face of the detector holder contains only a thick W1 detector.
The angular resolution of the W1-type silicon detectors (and, by extension, the silicon detector telescopes) with respect to the point of particle emission is 3--5\textdegree~in the present detector geometry.
By employing energy loss tabulations provided by the Stopping and Range of Ions in Matter (SRIM) software \cite{Zie10}, the angular resolution allows for high-resolution energy reconstruction of the initial proton energies, correcting for the energy losses endured by the protons in the catcher foil and the dead layers of the silicon detectors.
Table \ref{tab:silicons} lists the active detector thicknesses, the low-energy thresholds of the detectors and their solid angle coverages.

\begin{figure}
\includegraphics[width=0.98\columnwidth]{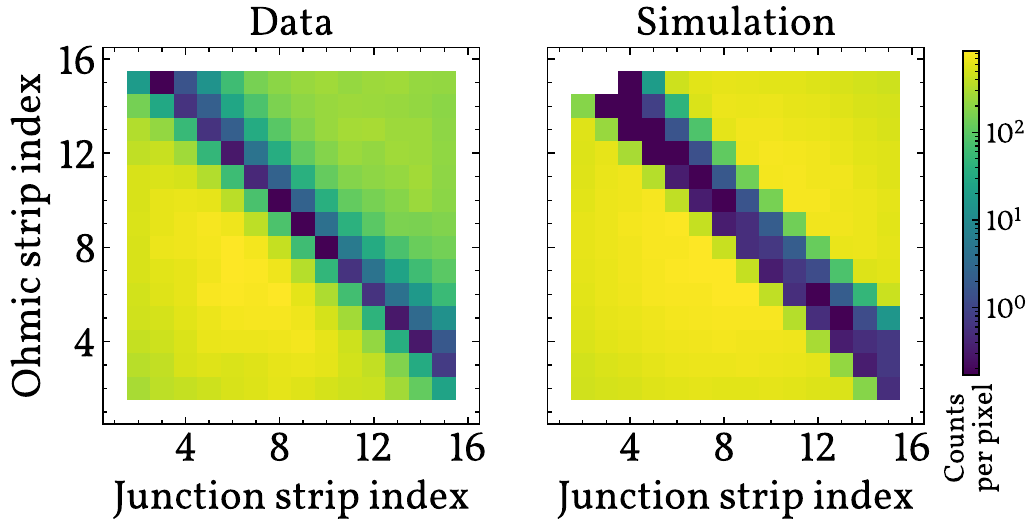}
\caption{\label{fig:shadow}%
Hit patterns for $\Delta E$-6, positioned below the catcher foil.
The shadow in the hit patterns are caused by the frame of the catcher foil.
\textbf{Left:} Distribution from real $\beta$-delayed protons emitted in the decay of \sili{}.
\textbf{Right:} Distribution based on simulations of isotropically emitted protons from a fixed point in the catcher foil.
}
\end{figure}

Hit patterns for $\Delta E$-6, positioned below the catcher foil, are shown in Fig. \ref{fig:shadow}.
In the figure, hit patterns are drawn based both on actual \sili{} data as well as simulated mono-energetic protons emitted isotropically from a fixed point in the catcher foil.
The shadow caused by the frame of the catcher foil is evident in both hit patterns.
The fixed point in the catcher foil was found by optimizing the simulated hit pattern to best reproduce that drawn with real data; see \cite{Jens24} for more details.
In the real data, the variation in intensity upstream compared to downstream of the shadow is attributed to the finite size of the beam spot in combination with the catcher foil not being stretched completely taut across its frame, the latter resulting in some uncertainty, of order 0.1 mm, in the position of the implantation spot parallel to the beam axis.
The same type of simulation was used to determine all solid angle coverages listed in Table \ref{tab:silicons}.

Neglecting the shadow caused by the frame of the catcher foil in the top and bottom faces of the detector holder, all $\Delta E$ detectors have solid angle coverages of around 8.0--8.7\% of $4\pi$ with respect to the beam implantation spot.
The backing $E$ detectors, on the other hand, have solid angle coverages of around 7.0--7.7\% of $4\pi$.
Charged particles emitted in the decay of \sili{} could, then, conceivably punch through the outer strips of the $\Delta E$ detectors and miss the corresponding backing $E$ detectors.
Generally speaking, it is not possible to differentiate between particles stopped in the outer strips of the $\Delta E$ detectors and particles that punch through the outer strips of the $\Delta E$ detectors but are never stopped in the backing $E$ detectors.
This effect and its implications are discussed at length in \cite{Jen23}.
In the present experiment, some of the protons emitted in the decay of \sili{} are energetic enough to punch through most of the $\Delta E$ detectors, and events in the outermost strips of these detectors are hence discarded.
This limits the solid angle coverage of the $\Delta E$-$E$ telescopes to a bit less than those of the backing $E$ detectors, as is reflected in Table \ref{tab:silicons}.

The preamplified signals from all detectors were provided as inputs to Mesytec MSCF-16 shaping amplifiers, the outputs of which were provided as inputs to Mesytec MADC-32 analog-to-digital converters.
The junction and ohmic sides of each DSSD had a dedicated trigger in their respective shaping amplifier, and a logical 'AND' was formed between the junction and ohmic side triggers of each DSSD.
Readout of the DAQ system was triggered by a logical 'OR' between the six DSSDs, the five SSDs and the two germanium detectors.
The trigger logic was handled by a GSI Vulom4b VMEbus board, and the data acquisition software was running on a Motorola MVME5500 VMEbus Single-Board Computer.

Further details on the employed detectors are given in appendix \ref{app:detector-details}.
The following two subsections describe the calibrations of the germanium detectors and the silicon detectors, in turn.

\subsection{Calibration of germanium detectors}

The two germanium detectors were energy-calibrated with a $^{152}$Eu source and with a $^{226}$Ra source to well-known $\gamma$-lines.
To determine the peak positions of the relevant $\gamma$ lines, the peak shapes were fitted with Exponentially Modified Gaussian (EMG) distributions to account for the asymmetric low-energy tail associated with detector response effects such as incomplete charge collection.
The EMG parametrization follows that of \cite{Pur17},

\begin{equation}
\begin{split}
f(x;N,\mu,\sigma,\tau) = &\,\frac{N}{2\tau}\exp{\left(\left(\frac{\sigma}{\sqrt{2}\tau}\right)^2+\frac{x-\mu}{\tau}\right)}\\
                         & \times\mathrm{erfc}\left(\frac{\sigma}{\sqrt{2}\tau}+\frac{x-\mu}{\sqrt{2}\sigma}\right),
\end{split}
\label{eq:gamma-peak-shape}
\end{equation}
where $x$ is the independent variable, $N$ is the integral under the curve, $\mu$ is the peak position, $\sigma$ is a width parameter of the peak, $\tau$ is an exponential decay parameter and $\textrm{erfc}(\cdot)$ represents the complementary error function.
In the fits, a local linear background was added to the EMG distribution.
The resulting peak position values were fitted linearly to the corresponding $\gamma$ energies.

The two germanium detectors were efficiency-calibrated with the same $^{152}$Eu and $^{226}$Ra sources as were used for their energy calibration.
The integral under the curve, $N$, in Eq. (\ref{eq:gamma-peak-shape}) was used to determine the number of counts within each peak.
The energy-dependent efficiency, $\epsilon$, of the germanium detector G1 was fitted to the expression

\begin{equation}
\ln{\epsilon}=A\ln{E}+B\,\ln^{2}{E}-\frac{C}{E^{3}},
\label{eq:g1-eff}
\end{equation}
where $E$ is the relevant energy and $A$, $B$ and $C$ are fit parameters.
The energy-dependent efficiency of the germanium detector G2 was fitted to the expression

\begin{equation}
\epsilon = A\,E^{B},
\label{eq:g2-eff}
\end{equation}
where the symbols have the same meaning.
The use of different efficiency parameterizations is due to an observed stronger low-energy cutoff in G1 as compared to G2.
The resulting fit parameters of the efficiency calibrations are given in Table \ref{tab:gamma-eff}.

\begin{table}
\caption{\label{tab:gamma-eff}%
Detection efficiency models and fit parameters of the germanium detectors G1 and G2 for $\gamma$ energies in keV.
Statistical uncertainties on the fit parameters are given in parentheses.
}
\begin{ruledtabular}
\begin{tabular}{r|cccccc}
   & Model                 & $A$        & $B$                   & $C$ \\
\colrule
G1 & Eq. (\ref{eq:g1-eff}) & $-1.23(2)$ & $6.4(3)\times10^{-2}$ & $1.2(8)\times10^{7}$ \\
G2 & Eq. (\ref{eq:g2-eff}) & $0.138(4)$ & $-0.500(4)$           &
\end{tabular}
\end{ruledtabular}
\end{table}

As mentioned previously, the DAQ did not offer ideal shaping and signal processing times for the germanium detectors; as such, the efficiencies and energy resolutions presented here are not indicative of the performance of the detectors themselves.
The detection efficiencies of G1 and G2 range from about 0.5\% at 0.5 MeV down to 0.3\% at 2.5 MeV, while the energy resolutions range from 5 keV FWHM (Full-Width at Half-Maximum) up to 20 keV FWHM.
More details on the energy and efficiency calibrations of the germanium detectors can be found in \cite{Ede24}.

\subsection{Calibration of silicon detectors}

\subsubsection{Energy calibration}\label{subsubsec:silicon-cal}

\begin{figure}
\includegraphics[width=0.80\columnwidth]{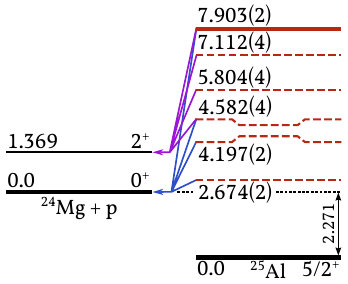}
\caption{\label{fig:25si-cal}%
Level diagram of $\beta$-delayed protons from \sili{} used for internal calibration of silicon detectors.
Energies are given in MeV relative to the ground states of \alum{} and \magn{}.
The uncertainties on the level energies in \alum{} are given in parentheses; they are orders of magnitude larger than both the uncertainties on the level energies in \magn{} and the \alum{} proton separation energy.
}
\end{figure}

\begin{table}
\caption{\label{tab:25si-cal}%
Kinetic energies of $\beta$-delayed protons, $E_p$, used for internal calibration of silicon detectors.
The protons are emitted from excited states in \alum{} with center-of-mass energies $E_i$ to states in \magn{} with center-of-mass energies $E_f$.
The proton separation energy of \alum{} is $S_p = 2271.37(7)$ keV.
The right-most column lists the detectors, with the same numbering scheme as in Table \ref{tab:silicons}, which used the various kinetic energies, $E_p$, for calibration.
The SSD $E$-4 was energy-calibrated with $\alpha$ particles and adjusted for pulse height defect empirically; see text.
}
\begin{ruledtabular}
\begin{tabular}{l|cccc}
 &$E_i$ (keV) & $E_f$ (keV) & $E_p$ (keV) & Used to calibrate \\
\colrule
$p_1$ & 2674(2) & 0           &  386(2) & $\Delta E$-1,2,3,4,6\\
$p_3$ & 4197(2) & 0           & 1848(2) & $\Delta E$-1,2,3,4,5,6\\
$p_2$ & 4582(4) & 1368.667(5) &  904(4) & $\Delta E$-1,2,3,4,5,6\\
$p_5$ &         & 0           & 2217(4) & $\Delta E$-1,2,3,4,5,6\\
$p_4$ & 5804(4) & 1368.667(5) & 2077(4) & $\Delta E$-1,2,3,4,5,6\\
$p_6$ & 7112(4) & 1368.667(5) & 3332(4) & $\Delta E$-4,5; $E$-1,2,3,6\\
$p_7$ & 7903(2) & 1368.667(5) & 4091(2) & $\Delta E$-4,5; $E$-1,2,3,6\\
$p_8$ &         & 0           & 5405(2) & $\Delta E$-4,5; $E$-1,2,3,6
\end{tabular}
\end{ruledtabular}
\end{table}

The silicon detectors were energy-calibrated internally to well-determined excited states in \alum{} which are populated in the $\beta$-decay of \sili{} and depopulated via proton emission; see decay scheme relevant for the calibrations in Fig. \ref{fig:25si-cal} and a list of calibration energies in Table \ref{tab:25si-cal}.
The employed literature excited states in \alum{} are known to high precision and accuracy from $^{24}\textrm{Mg}(\textrm{p},\gamma)$ and $^{24}\textrm{Mg}(\textrm{p},\textrm{p}' \gamma)$ reaction studies \cite{Bas25}.
The excited states in \alum{} are depopulated via proton emission primarily to the ground and first excited states of \magn{}; both the proton separation energy of \alum{} and the excitation energy of the first excited state in \magn{} are known to high accuracy and precision, as well \cite{Hua21,Wan21,Bas22}.
Ultimately, the limitation in achievable precision from the internal energy calibration in the present experiment is instrumental and is due to uncertainties in the peak fitting, the scatter among telescopes and uncertainties in energy-loss-corrected energies; the achievable energy precision is around 3 keV.

The initial proton kinetic energies, $E_p$, used for calibration are simply the differences in initial state energies, $E_i$, of \alum{} and final state energies, $E_f$, of \magn{} in the center-of-mass (\alum{} is assumed to break up at rest), less the proton separation energy, $S_p$, of \alum{} and moderated by the relevant mass fraction,

\begin{equation}
E_p = \frac{M}{M+m}(E_i - E_f - S_p),
\label{eq:proton-energy}
\end{equation}
where the proton separation energy, $S_p$, the proton mass, $m$, and the mass of \magn{}, $M$, are adopted from the Atomic Mass Evaluation (AME) 2020 \cite{Hua21,Wan21}.
The energies detected in the different silicon detector segments are not the initial proton kinetic energies, $E_p$, but the deposited energies, $E_{\textrm{dep}}$, in the active detector segments, following energy losses in traversing the catcher foil and the dead layers of the detector segments.
For each detector segment in the DSSDs, the solid angle-averaged deposited energies $\overline{E_{\textrm{dep}}}(E_p,\overline{\theta})$, based on the initial proton energies, $E_p$, and the solid angle-averaged angle of incidence with respect to normal between proton and detector segment, $\overline{\theta}$, were calculated using SRIM; $\overline{\theta}$ being used to determine the effective thicknesses of traversed inactive media.
These energies were then used for calibration.

\begin{figure}
\includegraphics[width=1.0\columnwidth]{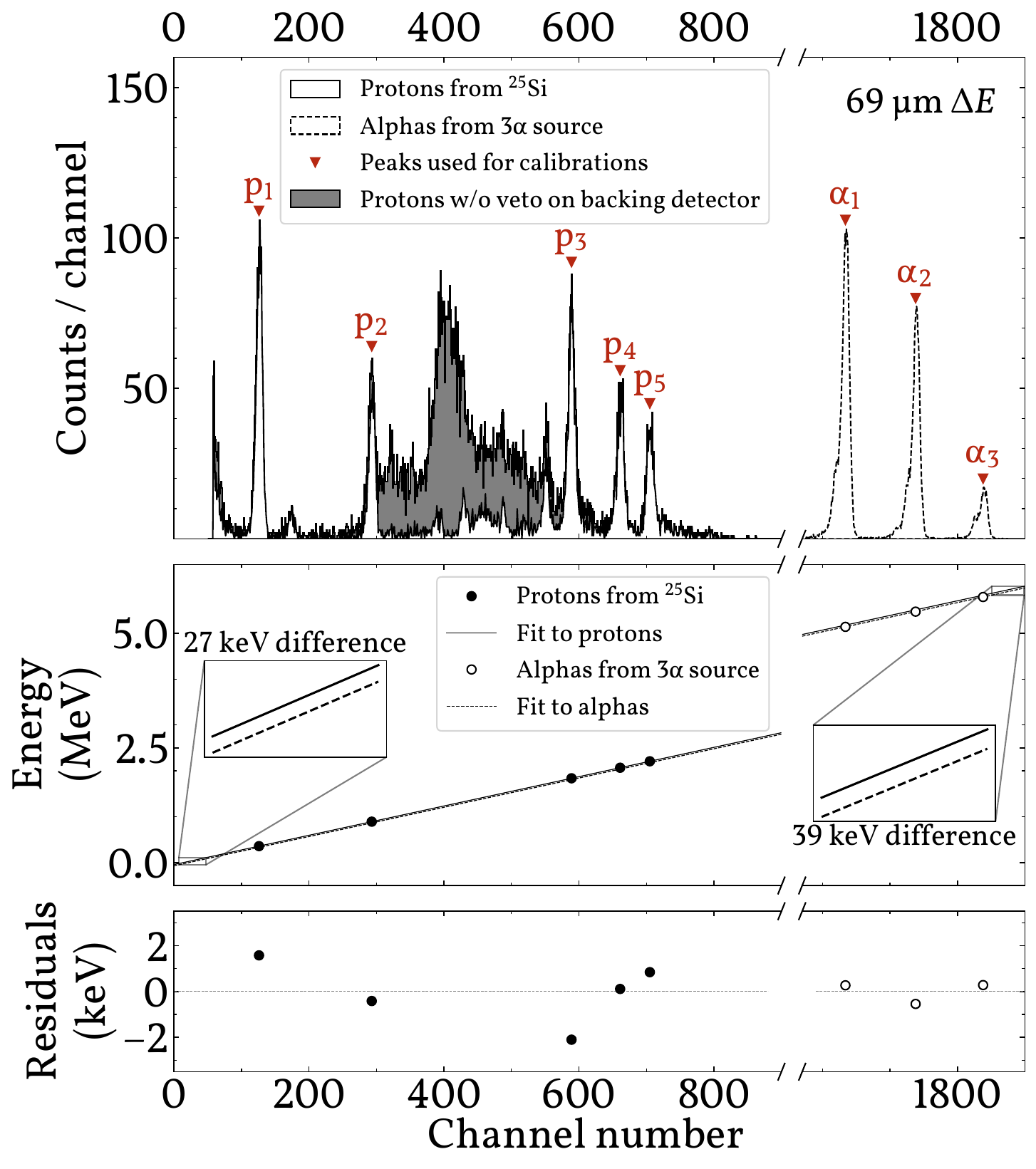}
\caption{\label{fig:silicon-strip-cal}%
Energy calibration of a central junction-side strip of thin DSSD $\Delta E$-2 to prominent proton peaks from the decay of \sili{} at energies below the punch through energy of the strip.
\textbf{Top:} Identification of proton peaks and, for comparison, identification of $\alpha$ particle peaks from a standard triple-$\alpha$ source.
Punch through events are removed from the proton spectrum by imposing a veto on signals in the backing $E$ detector.
\textbf{Middle:} Calibration to the deposited energies $E_{\textrm{dep}}$ (see text) of both protons and $\alpha$ particles.
\textbf{Bottom:} Residuals of the fit of deposited energies to the corresponding channel values.
}
\end{figure}

Below the punch through energies of the $\Delta E$ detectors, the singles energy spectra of the $\Delta E$ detectors will contain punch through events.
In order to clear the energy spectra of punch through events, a veto on signals in the backing $E$ detectors was imposed when locating proton peaks below punch through energies.
In general, the peak positions of the relevant proton lines were found using an automatic peak-identification algorithm described in \cite{Mar67}.
The resulting peak position values were fitted linearly to the corresponding deposited energies.
An example spectrum for a thin DSSD detector, $\Delta E$-2, with identified peak positions below proton punch through energy is shown in Fig. \ref{fig:silicon-strip-cal}.
The figure also illustrates identification of peaks from a standard triple-$\alpha$ source and the difference in calibrations to the proton peaks and the $\alpha$ peaks.
The extrapolations of the $\alpha$ calibration to the proton energy range and vice-versa are rather large.
Still, the well-known pulse height defect seems to be responsible for a slight offset in absolute energies as well as a small relative difference over the full range, as expected.

\begin{figure}
\includegraphics[width=1.0\columnwidth]{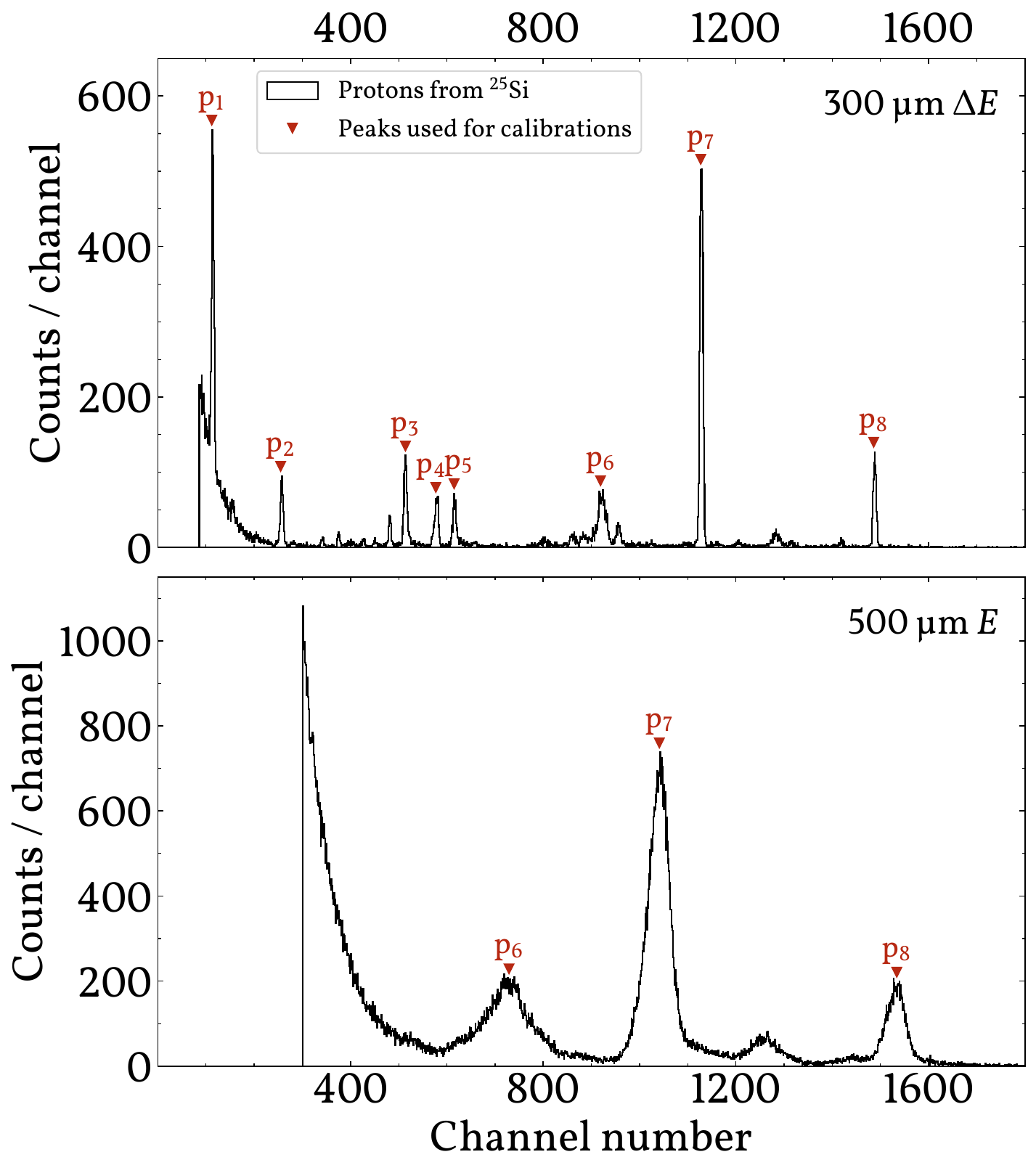}
\caption{\label{fig:silicon-thick-dssd-pad-cal}%
Peak identification of reference proton energies in thick silicon detectors.
\textbf{Top:} Spectrum from central junction-side strip of DSSD $\Delta E$-4 (300 µm).
\textbf{Bottom:} Spectrum from SSD $E$-3 (505 µm), which detects only protons energetic enough to punch through the DSSD $\Delta E$-3 in front of it.
The difference in peak intensities of $p_1$ in the top panel of this figure compared to Fig. \ref{fig:silicon-strip-cal} is due to the peak's vicinity to detector channel thresholds; see text.
}
\end{figure}

The thick DSSDs have higher punch through thresholds and, hence, reference protons with larger kinetic energies are fully stopped in them, as exemplified in the top panel of Fig. \ref{fig:silicon-thick-dssd-pad-cal}, where all reference energies in Table \ref{tab:25si-cal} are identified in the 300 µm DSSD, $\Delta E$-4.
For the thickest DSSD, $\Delta E$-5, with an active thickness of around 1000 µm, the $\beta$ response at low energy dominates the reference proton of lowest energy ($p_1$ in the top panels of Figs. \ref{fig:silicon-strip-cal} and \ref{fig:silicon-thick-dssd-pad-cal}) and this potential calibration point has to be discarded.

The low-energy peak $p_1$ in the top panels of Figs. \ref{fig:silicon-strip-cal} and \ref{fig:silicon-thick-dssd-pad-cal} has vastly different intensities in the two detector channels, as compared to the other peaks shared between the two spectra.
The low-energy thresholds were set individually for each detector channel, and the lowest proton peak at an energy $E_p =$ 386(2) keV turns out to be placed in the threshold region so that its detection efficiency varies across all detectors.
This hinders the possibility of extracting proper intensities for this specific peak in the analysis of the $\beta$-delayed protons in section \ref{subsubsec:beta-p}.
Protons at higher energies do not have this issue.

The backing SSDs of the setup are calibrated, to the extent possible, with reference proton energies from protons that punch through the DSSDs in front of them.
In order to be able to determine the energy deposition in the SSDs to high accuracy in this manner, accurate knowledge of the active and inactive layer thicknesses of the preceding DSSDs is needed; methods employed to determine DSSD thicknesses are described in appendix \ref{app:thickness-estimation}.
Energy losses in all media preceding the active layers of the SSDs are taken into account in line with the description above of evaluating $\overline{E_{\textrm{dep}}}(E_p,\overline{\theta})$ for the DSSDs.
However, accurate information on $\overline{\theta}$ from punch through of the DSSDs was not taken into account, as proper identification of peaks in smaller subsets of events from the SSDs in various ranges of $\overline{\theta}$ proved difficult with the available statistics.
We believe, however, that there is the potential to achieve energy resolution and accuracy in the backing SSDs approaching the same levels achieved in the preceding DSSDs, given sufficient statistics to allow the generation of pulse height spectra of the backing SSDs gated on subsets of punch through events of the relevant DSSD, ideally on a pixel-by-pixel basis (this would of course also benefit the calibrations of the individual DSSD channels).
With the ongoing increase in beam power at FRIB as well as improvements in beam transmission at the facility, the yields of \sili{} achievable in the Gas Stoppig Area could reach a point where this is feasible, even for a relatively brief calibration run with \sili{}.
An example of peak identification of reference protons in an SSD is shown in the bottom panel of Fig. \ref{fig:silicon-thick-dssd-pad-cal}.

Preceding the SSD $E$-4 was the DSSD $\Delta E$-4 of 300 µm active layer thickness.
As all reference protons stopped in $\Delta E$-4, an absolute energy scale for $E$-4 could only be achieved with a separate calibration measurement using a standard triple-$\alpha$ source.
The pulse height defect, responsible in this case for a systematic offset in observed proton energies when calibrating with known $\alpha$ particle kinetic energies, is apparent in the middle panel of Fig. \ref{fig:silicon-strip-cal}, for the example of a thin DSSD strip.
Separate $\alpha$ calibration measurements were carried out for all SSDs.
The relative differences in calibration coefficients achieved when calibrating $E$-1, $E$-2 and $E$-3 with either protons or $\alpha$ particles were compared:
The mean relative difference in the slope of the linear relation between observed energy and channel value was found to be $a_{p/\alpha} =$ 1.014(1) and the mean relative difference in the offset was found to be $b_{p/\alpha} =$ 1(1) keV, consistent with other studies of the pulse height defect for low-energy protons and $\alpha$ particles; see e.g. \cite{Len86}.
The energy calibration of $E$-4 to standard triple-$\alpha$ source calibration peaks was adjusted accordingly with the parameters $a_{p/\alpha}$ and $b_{p/\alpha}$.
Further details on the calibration of the silicon detectors can be found in \cite{Jens24,Ped25}.

\subsubsection{Reconstruction of proton energy spectra}\label{subsubsec:reconstruction}

Detector telescopes are widely used for particle/nuclide identification based on the energy-loss signatures in $\Delta E$ and $E$ detectors, when the incident particles/nuclides are sufficiently energetic to punch through $\Delta E$ detectors and be completely stopped in backing $E$ detectors.
In low-mass, proton-rich $\beta$-decay, the types of charged particles emitted following $\beta$-decay are limited to protons and $\alpha$ particles, and it is hence possible—with a bit of care and reference to known nuclear structure of the proton- and $\alpha$-daughters—to identify protons below the punch through energies of the thin (thickness less than 100 µm) $\Delta E$ detectors employed in a setup such as the one of the present experiment.
The benefit of employing \mbox{$\Delta E$-$E$} detector telescopes in low-mass, proton-rich $\beta$-decay experiments is the limited response of the thin $\Delta E$ detectors to the $\beta$ particles, enabling background-free $\beta$-delayed proton spectra towards low energy, as demonstrated e.g. in Fig. \ref{fig:silicon-strip-cal} where the peak $p_1$ originates from a proton emitted with energy \mbox{$E_p =$ 386(2) keV} and the tail of the $\beta$ response is seen at energy slightly below it.
In the following, we shall consider only the reconstruction of initial kinetic energies of protons below punch through energies of $\Delta E$ detectors, but the arguments extend, in principle, to any heavier type of charged particle.

\begin{figure}
\includegraphics[width=0.98\columnwidth]{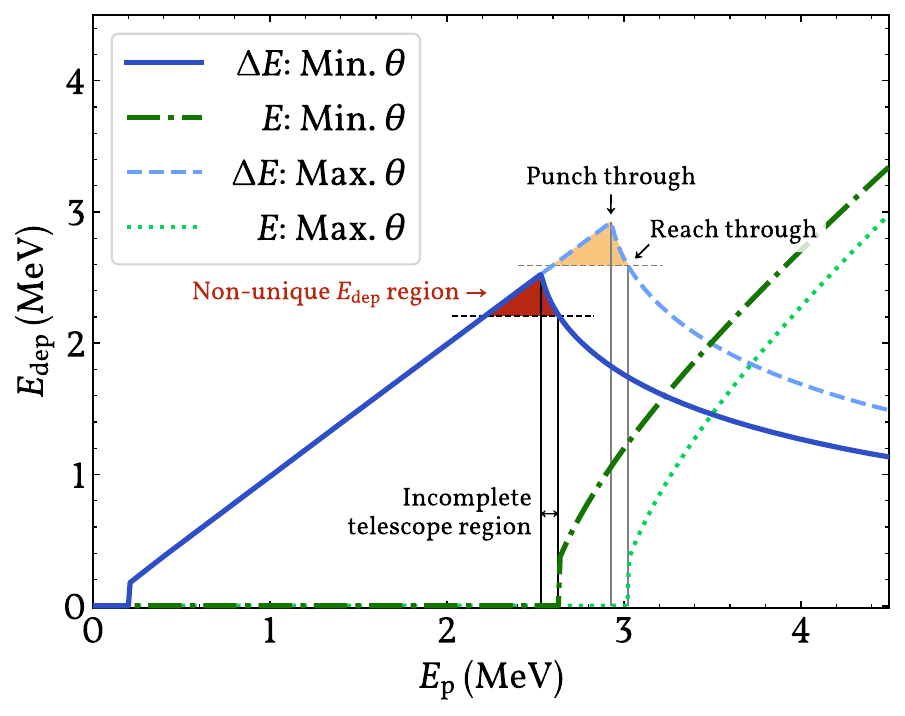}
\caption{\label{fig:telescope-spurious}%
Deposited energy, $E_{\mathrm{dep}}$, vs. initial proton kinetic energy, $E_p$, in $\Delta E$ and $E$ detector at small and large angles of incidence with respect to detector surface normal, $\theta$.
The example is for \mbox{$\Delta E$-$E$} detector telescope no. 2 with $\Delta E$ active thickness 69 µm and with low-energy thresholds $\sim$200 keV in the $\Delta E$ detector and 340 keV in the $E$ detector.
The low-energy thresholds effectively add to the detectors' dead layers.
The minimum value of $\theta$ is 0\textdegree, and the maximum value is 35\textdegree.
}
\end{figure}

When extending the reconstructed initial proton kinetic energy spectra, $E_p$, below the punch through energies of the $\Delta E$ detectors, the \emph{incomplete telescope region} of each \mbox{$\Delta E$-$E$} detector telescope induce telescope-specific regions in the observed deposited energies, $E_{\mathrm{dep}}$, where $E_{\mathrm{dep}}$ is \emph{not} unique, thus making the reconstruction of $E_p$ dubious in the non-unique region of $E_{\mathrm{dep}}$.
This effect is discussed at length in \cite{Jen23}, and the main concepts of the paper, encapsulated in its Fig. 1, are redrawn here, in Fig. \ref{fig:telescope-spurious}, for the case of $\Delta E$-$E$ telescope no. 2 of the present experiment.
The figure is drawn by employing energy loss tables from SRIM.

The incomplete telescope region of a $\Delta E$-$E$ detector telescope is the interval of initial proton kinetic energies, $E_p$, where the proton is energetic enough to punch through the $\Delta E$ detector but is not sufficiently energetic to \emph{reach through} to the $E$ detector.
The dead layers between the active layers of the $\Delta E$ and the $E$ detectors as well as the low-energy threshold of the $E$ detector dictates the extent of the detector telescope's incomplete region.
As is shown in Fig. \ref{fig:telescope-spurious}, the energy deposited by a proton in the $\Delta E$ detector increases linearly (after the low-energy threshold of the $\Delta E$ detector has been overcome) with the initial proton kinetic energy until the punch through threshold—the maximally possible energy deposition in the $\Delta E$ detector—occurs.
The deposited energy in the $\Delta E$ detector then decreases non-linearly until the proton reaches through to the backing $E$ detector and the energy deposition in the $E$ detector becomes finite.
As the effective thickness of the $\Delta E$ detector scales as $1/\cos{\theta}$, where $\theta$ is the angle of incidence with respect to detector surface normal, the incomplete telescope region is inherently $\theta$-dependent.
The bidirectional arrow in Fig. \ref{fig:telescope-spurious} indicates the extent of the incomplete telescope region for $\theta =$ 0\textdegree~for $\Delta E$-$E$ detector telescope no. 2 (Table \ref{tab:silicons}), while punch through and reach through is indicated for $\theta =$ 35\textdegree~for the same detector telescope.

The triangular filled regions, respectively for \mbox{$\theta =$ 0\textdegree} and for $\theta =$ 35\textdegree, with energy depositions between punch through and reach through in Fig. \ref{fig:telescope-spurious}, indicate regions where the energy depositions, at given $\theta$, are not unique.
Without any knowledge informing whether an observed energy, $E_{\mathrm{dep}}$, in these non-unique regions should belong to the high-energy or the low-energy end of the ranges of initial proton kinetic energies, $E_p$, events in these regions must, then, be discarded.
Note that this discarding of events introduces an energy- and $\theta$-dependence in the solid angle coverage of a given detector telescope; i.e. the solid angle coverages in Table \ref{tab:silicons} in the present experiment have to be corrected for this effect.

\begin{figure}
\includegraphics[width=1.00\columnwidth]{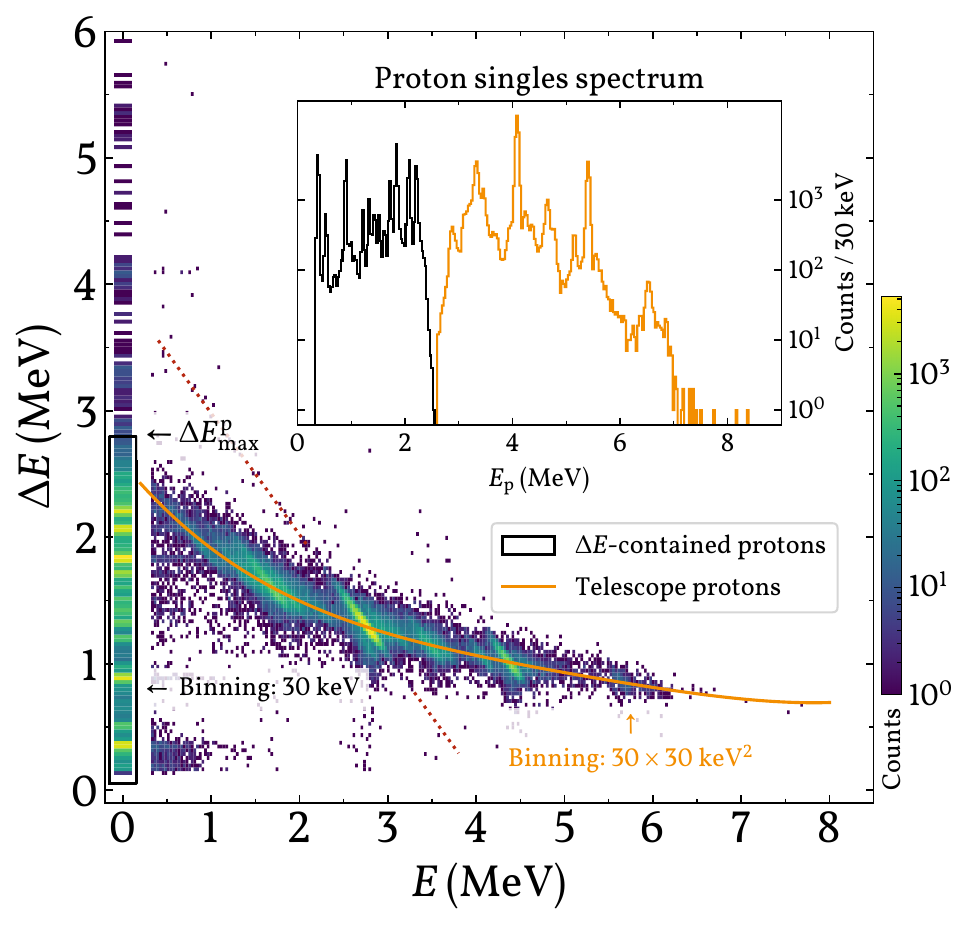}
\caption{\label{fig:banana-U2}%
Energy deposition in the $\Delta E$ detector vs. energy deposition in the $E$ detector of the $\Delta E$-$E$ detector telescope no. 2 of the setup.
Sufficiently energetic protons punch through the $\Delta E$ detector and are stopped in the $E$ detector, generating \emph{telescope proton events}.
Events with energy deposition only in the $\Delta E$ detector without any corresponding energy deposition in the $E$ detector, generating \emph{$\Delta E$-contained proton events}, are drawn with a width of 200 keV along the $E$ axis, centered at $E=0$.
The indicated $\Delta E_{\max}^{p}$ is the maximum energy a proton can deposit in the $\Delta E$ detector.
The inset of the figure shows the reconstructed proton energies, $E_p$, from both the $\Delta E$-contained proton events and the telescope proton events.
The region in the inset where the number of $\Delta E$-contained events go to zero from below and the number of telescope events go to zero from above is instrumental: It is due to the inherent \emph{incomplete region} of the detector telescope; see text.
The dotted red line is meant to guide the eye; see text for discussion of the events near it above $\Delta E \sim$ 2 MeV.
}
\end{figure}

Finally, after identifying the dubious energy regions below punch through threshold of the $\Delta E$-$E$ detector telescopes, complete proton energy spectra based on the observed energies can be reconstructed, where the dubious energy regions are excluded.
A \mbox{$\Delta E$ vs. $E$} plot is shown in Fig. \ref{fig:banana-U2} for the example of $\Delta E$-$E$ detector telescope no. 2 (Table \ref{tab:silicons}).
In the figure, protons with energies below punch through threshold (roughly indicated with $\Delta E_{\max}^{p}$) are completely stopped in the $\Delta E$ detector (\emph{$\Delta E$-contained protons}), and protons that manage to reach through to the backing $E$ detector deposit a fraction of their energy in the $\Delta E$ detector before being completely stopped in the $E$ detector (\emph{telescope protons}).
The inset of the figure shows the reconstructed proton energy spectrum from the two types of detection events; the region where the number of $\Delta E$-contained events go to zero from below and the number of telescope events go to zero from above is due to the incomplete region of the detector telescope.

The $\Delta E$-$E$ spectra of the detector telescopes accumulated throughout the entire experiment have revealed that not all of the beam was stopped in the catcher foil.
Rather, a small fraction of the beam managed to enter the silicon detector holder and implant on a surface from which protons with very large angles of incidence with respect to the silicon detector telescopes could be emitted.
This explains the small accumulation of events near the red dotted line in Fig. \ref{fig:banana-U2}.
The effect is most clear for the most intense corresponding proton transition energies in the $\Delta E$-$E$ spectra.

While this effect does not influence the reconstructed proton energy spectra, which are only composed of the $\Delta E$-contained events and telescope events (inside the ``banana''), it does bring into question the absolute efficiency of detection of \sili{} stopped in the catcher foil of the setup:
It is not possible to differentiate in the $\gamma$ spectra of the germanium detectors between beam stopped in the foil and beam implanted elsewhere inside the silicon detector holder.
Furthermore, if there is yet another fraction of the beam implanted where $\beta$-delayed protons cannot reach the silicon detectors, but $\gamma$-rays from the decay are still detected in the germanium detectors, this adds an additional uncertainty to the difference between detection efficiency of the silicon and germanium detectors.

The events above proton punch through threshold in the $\Delta E$ detector in Fig. \ref{fig:banana-U2} is a low-intensity background of long-lived $\alpha$ calibration sources implanted prior to the experiment.
No $\alpha$ particles are expected to be emitted in the decay of \sili{} (see Fig. \ref{fig:decay-scheme-overview}), but, for cases where the $\alpha$ particle kinetic energies are greater than the proton punch through thresholds of the thin $\Delta E$ detectors, unambiguous identification of $\alpha$ particles is possible (in the present mass region) since protons and $\alpha$ particles are the only charged particles emitted following $\beta$-decay here.

\section{Results on the decay of \sili{}}

\begin{figure}
\includegraphics[width=0.98\columnwidth]{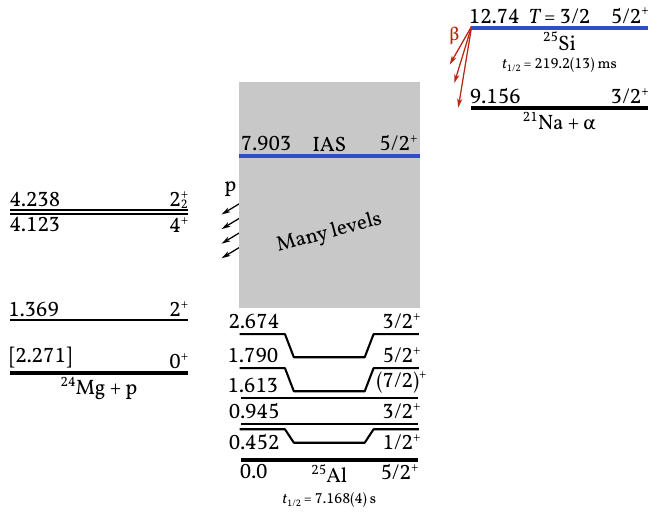}
\caption{\label{fig:decay-scheme-overview}%
Overview of the decay scheme of \sili{}.
Energies are given in MeV relative to the ground state of \alum{}, except for excited states in \magn{} which are given relative to the ground state of \magn{}.
Not all individual levels in \alum{} that take part in proton decay are shown explicitly; they are shown in Fig. \ref{fig:decayscheme} and listed in Table \ref{tab:protons}.
Masses, separation energies and other level parameters adopted from \cite{Bas25,Bas22,Hua21,Wan21}.
}
\end{figure}

An overview of the decay scheme of \sili{} is presented in Fig. \ref{fig:decay-scheme-overview}.
The main $\gamma$-lines observed in the decay are from states below and close to the proton separation energy, $S_p = 2271.37(7)$ keV \cite{Hua21,Wan21}, in \alum{} and from de-excitation of the $4^+$ and $2^+$ levels in \magn{} \cite{Sun21}.
Protons are dominantly emitted from excited states in \alum{} above the proton separation energy to states in \magn{} -- in particular from the Isobaric Analogue State (IAS).
99\% of the $\beta$-decay of the daughter nuclide \alum{} is to the ground state of the stable nuclide $^{25}$Mg \cite{Bas25}.
The $\alpha$ separation energy of \alum{}, $S_{\alpha} = 9156.03(8)$ keV, is too large for the $\beta$-unstable $^{21}$Na to participate significantly in the decay of \sili{}.

\subsection{The structure of the entering states}\label{subsec:structure}

The mirror nuclei \alum{} and $^{25}$Mg constitute a classic example of deformation in light nuclei \cite{Boh75}.
The decay of \sili{} is quite similar to the one of $^{21}$Mg \cite{Jen24}; the nuclei that enter differ by an $\alpha$ particle, where the deformation of the states plays a significant role \cite{Rii24}, and we therefore briefly recount what is known in the present case.

\begin{table}
\caption{\label{tab:peaks}%
Levels in $^{25}$Al possibly fed in allowed $\beta$-decay of $^{25}$Si.
The literature values \protect\cite{Bas25} of excitation energies, $E_{\mathrm{ex}}$, widths, $\Gamma$, and spin-parities, $J^\pi$, are listed on the left along with band assignments \protect\cite{Fuj04}.
On the right, the proton center-of-mass energies relative to the $0^+$ ground state of \magn{}, $E_{p,0}^{\mathrm{cm}}$, from literature are compared with those extracted from the present experiment.
The literature values of $E_{p,0}^{\mathrm{cm}}$ are calculated based on level energy differences while the experimental values are calculated based on the observed proton kinetic energies.
}
\begin{ruledtabular}
\begin{tabular}{ccll|cl}
  $E_{\mathrm{ex}}$ (keV) & $\Gamma$ (keV) & $J^{\pi}$ & Band$^c$ & \multicolumn{2}{c}{$E_{p,0}^{\mathrm{cm}}$ (keV)} \\
  & & & & Lit. & Exp. \\
\colrule
  0 & 0 & $5/2^+$ & 1 & & \\
  945 & $^a$ & $3/2^+$ & 2 & & \\
  1613 & $^a$ & $(7/2)^+$ & 1 & & \\
  1790 & $^a$ & $5/2^+$ & 2 & & \\
  2674 & $^a$ & $3/2^+$ & 3 & 402.2(4)$^d$ & 401(3) \\
  2720 & $^a$ & $7/2^+$ & 2 & 449 & -- \\
  3859 & $^a$ & $5/2^+$ & 3 &  1587.7(8) & 1585(4) \\
  4197(2) & $^a$ & $3/2^+$ & & 1926(2)$^d$ & 1922(4) \\
  4582(2) & $^a$ & $5/2^+$ & & 2311(2)$^d$ & 2310(4) \\
  4906(4) &  & $(7/2)^+$ & 3 & 2635(4) & 2631(6) \\
  5597(5) & 56(20) & $^b$ & & 3326(5) & $^e$ \\
  5804(4) &  & $^b$ & & 3533(4) & 3531(4) \\
  6121(3) & 53(2) & $3/2^+$ & 4 & 3850(3) & 3845(8) \\
  6170(2) &  & $^b$ & & 4056(2) & $^e$ \\
  6518(7) & 59(13) & $3/2^+$ & & 4247(7) & -- \\
  6620(9) &  & $^b$ & & 4524(9) & $^e$ \\
  6649(5) & 58(9) & $5/2^+$ & 4 & 4378(5) & 4383(10) \\
  6879(7) &  & $^b$ & & 4608(7) & 4604(6) \\
  6944(10) & 104(10) & & & 4673(10) & 4694(15) \\
  7112(4) & 117(4) & $3/2^+$ & & 4841(4) & 4844(7) \\
  7242(3) & 19(4) & $5/2^+$ & & 4971(3) & 4973(5) \\
  7421(5) &  & $(7/2)^+$ & 4 & 5150(5) & 5154(5) \\
  7646 & 50(15) & $^b$ & & 5591 & $^e$  \\
  7716(10) & 230(20) & $3/2^+$ & & 5445(10) & 5450(50) \\
  7903(2) & $^a$ & $5/2^+$ & & 5632(2)$^d$ & 5632(5) \\
  7971(3) & 1.30(14) & $3/2^+$ & & 5700(3)  & -- \\
  8189(3) & 40(10) & $^b$ & &  5918(3) & 5921(4) \\
  9068(7) &  & $^b$ & &  6797(7) & 6805(10) \\
  9275(25) & & $^b$ & & 7004(25) & 7000(25) \\
  9415(30) & & $^b$ & & 7144(30) & $^e$ \\
\end{tabular}
\end{ruledtabular}
\begin{flushleft}
$^a$Width less than 1 keV.\\
$^b$$(3/2,5/2,7/2)^+$.\\
$^c$Single particle configurations on top of the $^{24}$Mg ground state: 1 = [202]5/2, 2 = [211]1/2, 3 = [200]1/2, 4 = [202]3/2. The notation $[N n_z \Lambda] \Omega$ is employed with $\Omega$ the component of total angular momentum along the symmetry axis, $N$ the total number of nodal surfaces for the single-particle wave function in a spheroidal Woods-Saxon potential with spin-orbit coupling, $n_z$ the number of nodal surfaces along the symmetry axis, and $\Lambda$ the component of orbital angular momentum along the symmetry axis.\\
$^d$Transition used for internal calibration of silicon detectors.
$^e$Level earlier reported fed in $\beta$-decay, but not confirmed here; see text.
\end{flushleft}
\end{table}

The band structure of \alum{} for the simple configurations of a proton on top of the ground state in \magn{} is now known up to 6--7 MeV \cite{Boh75,Fuj04}.
The levels in \alum{} that can be fed in allowed $\beta$-decay are presented in Table \ref{tab:peaks}; levels that belong to one of the four expected rotational bands are indicated.
All four bands are fed in $\beta$-decay.

The projection, $K$, of the total angular momentum along the symmetry axis in the deformed ground state of \sili{} is believed to be $K=3/2$, similarly to the ground state of its mirror nucleus $^{25}$Na (both have a very low-lying $3/2^+$ state, see e.g. \cite{Lon18}), but not with a pure rotational behavior, hence significant band mixing is believed to be present \cite{Ell68,Jan73,Col75}.
Meanwhile, the three lowest levels in \magn{} have $K=0$, followed by two levels with $K=2$.
Proton emission, where the proton carries total angular momentum $j$, from states in \alum{} with $K=K_i$ to states in \magn{} with $K=K_f$ that does not fulfill $|K_i - K_f| < j$ will be hindered \cite{Rii24}.
Significant differences in de-excitation intensities from the same initial state to different final states, where $K_f$ is known, can hint at the values of $K_i$ and/or $j$ involved in the decay, especially when single-particle penetrabilities alone cannot account for the differences \cite{Rii24,Jen24}.
We shall return to and expand upon this in section \ref{subsec:comments}.

\subsection{The $\beta$-transitions}

Since the structure of the nucleus \alum{} is well established, we shall follow \cite{Hat92} and start from the known levels in \alum{}.
As mentioned above, Table \ref{tab:peaks} lists all levels that could be fed in allowed $\beta$-decay from \sili{}; i.e.\ levels with spin-parity $3/2^+$, $5/2^+$ or $7/2^+$ as well as levels that have previously been reported to be fed in $\beta$-decay.
More details on the data analysis can be found in \cite{Ede24}.

\subsubsection{The $\beta\gamma$-decays}

A thorough investigation of the $\gamma$-ray spectrum resulting from the decay of $^{25}$Si was published recently \cite{Sun21}.
The investigation was conducted with higher detection efficiency, better energy resolution and larger energy range than our data and had more than an order of magnitude better statistics for singles $\gamma$-rays and much better $\gamma$-$\gamma$ coincidence rates.
The low-energy proton spectrum was also recorded.
This gave important proton-$\gamma$ coincidence information that will be employed later.
We shall therefore refer to \cite{Sun21} for a detailed discussion of the $\beta\gamma$-decay and mainly discuss the normalization of the different decay types here.

\begin{figure*}
\includegraphics[width=1.00\textwidth]{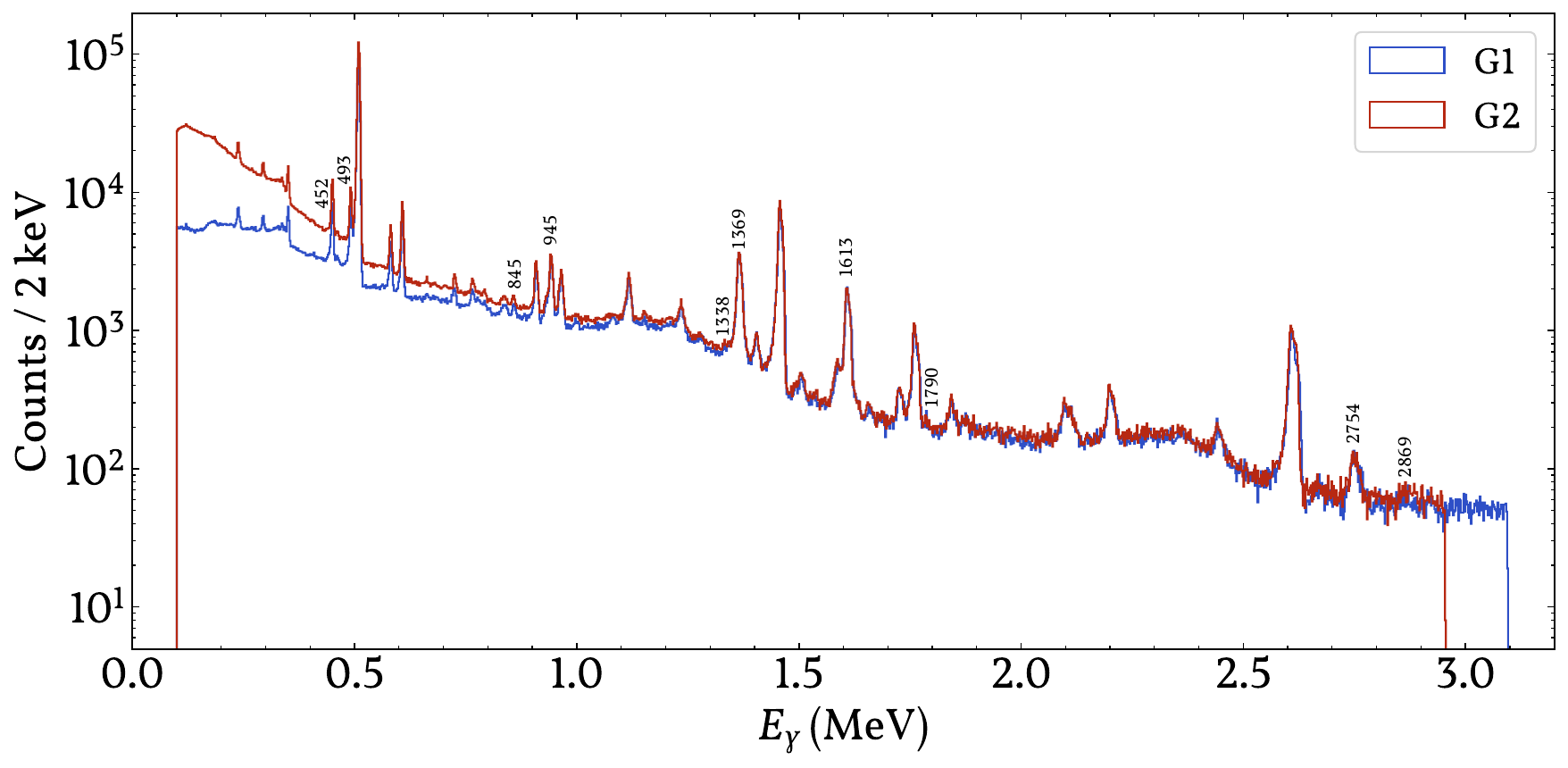}
\caption{\label{fig:gamma-spectra}%
$\gamma$ spectrum from the decay of \sili{} obtained with the two germanium detectors G1 and G2 (table \ref{tab:gamma-eff}).
The energies, $E_{\gamma}$, of the $\gamma$-lines relevant to the decay of \sili{} are labeled in keV; the transitions which these $\gamma$-lines correspond to are shown in Fig. \ref{fig:decayscheme}.
}
\end{figure*}

The $\gamma$ spectra obtained in the present experiment are shown in Fig. \ref{fig:gamma-spectra}.
The $\gamma$-lines relevant to the decay of \sili{} have their energies highlighted in keV.
The transitions which these $\gamma$-lines correspond to are shown in Fig. \ref{fig:decayscheme}.
In section \ref{sec:experiment}, the energy resolution and efficiency of the germanium detectors have already been discussed.
Again, we refer to \cite{Sun21} for a detailed study of the $\beta\gamma$-decay of \sili{}.

$\gamma$-decays are known to take place from several of the proton unbound levels in \alum{} and is the main decay route for the $7/2^+$ level at 2720 keV; see Fig. \ref{fig:decay-scheme-overview}.
No feeding of this level in $\beta$-decay has been reported so far.
It is assumed to belong to the lowest $1/2^+$ band and would then correspond to a core rotation of 4$\hbar$.
Proton emission from this state would be in a $g$-wave to the $0^+$ ground state of \magn{}; both $\beta$-feeding of this level as well as subsequent proton decay from it are highly suppressed.
Since only $\gamma$-decay has been observed from this level, we may extract a limit on the feeding from Fig. 4 in \cite{Sun21}:
The main line (75\% of the de-excitations) is at 1775.5 keV \cite{Bas25}, and an upper limit is clearly below the intensity of the observed 1789 keV line, giving a conservative upper value of 0.6\% per \sili{} decay.

We could not record the number of incoming ions and thus have no possibility of estimating the \alum{} ground state feeding experimentally.
In contrast, absolute normalizations were possible in the investigations carried out in \cite{Hat92,Tho04}.
In the former, the fraction of decays going to the four bound states was determined to be 58.7(15)\%, while in the latter, it was determined to be 65(2)\%.
These two values correspond to total proton branching ratios of 41.3(15)\% or 35(2)\%.
The ground state branching ratio was estimated theoretically to be 21.5(12)\% in \cite{Sun21}, which leads to a deduced fraction of bound state decays of 60.8(17)\%.
We shall employ a slightly conservative value of 61(3)\%.

The relative intensities of the $\gamma$-rays in $^{24}$Mg are reviewed in \cite{Sun21}.
All previous experiments are consistent and point to somewhat more protons being emitted to the $2^+$ state than to the $0^+$ ground state.
We shall return to this point below.

\subsubsection{The $\beta$p-decays}\label{subsubsec:beta-p}

The first $\beta$-decay experiments of $^{25}$Si did not have sufficient statistics to resolve all the features of the proton spectrum, but the main features were established early.
The spectra recorded in \cite{Rob93} and (for low proton energy) \cite{Sun21,Ste24} agree nicely with our results.

\begin{figure*}
\includegraphics[width=1.00\textwidth]{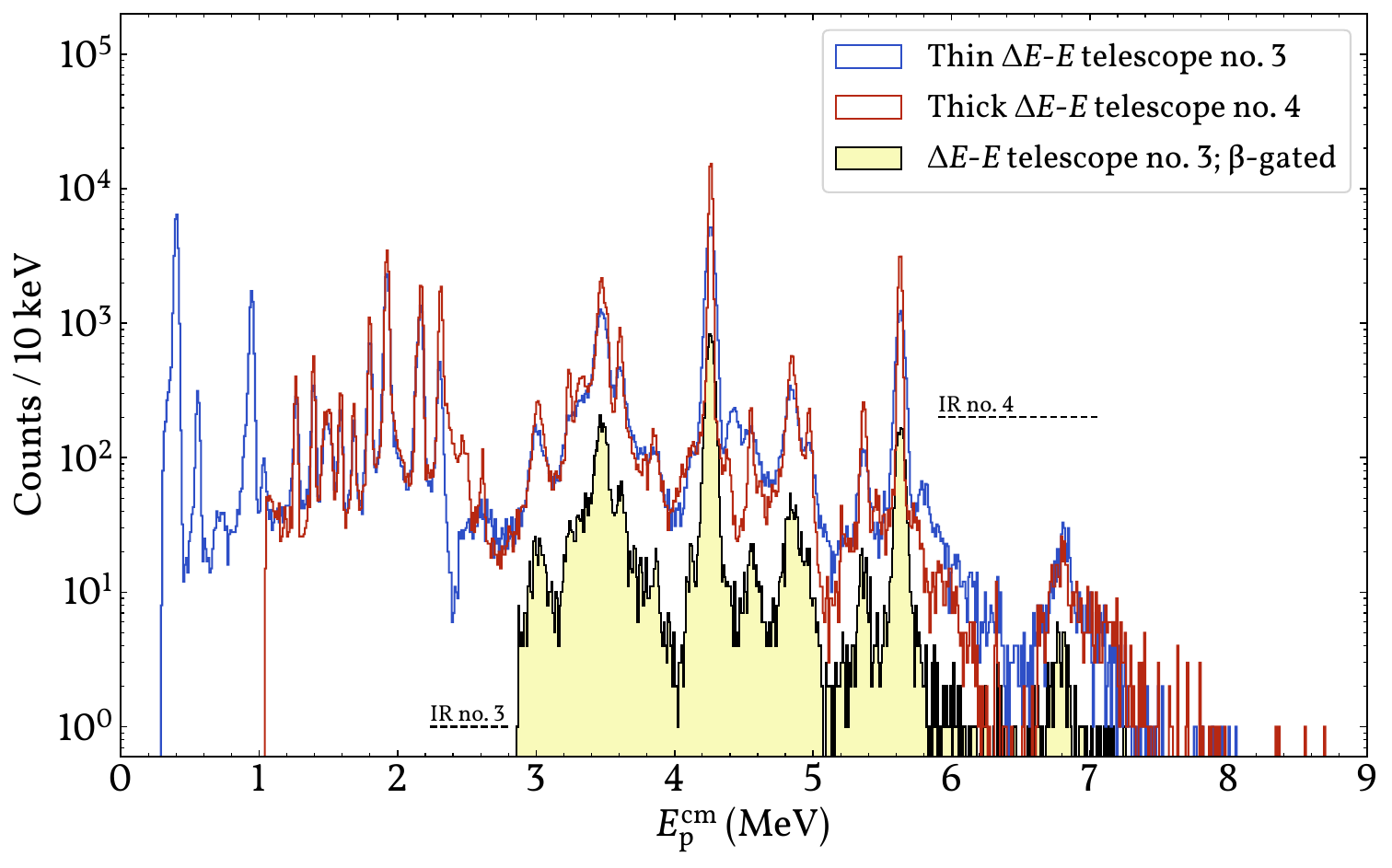}
\caption{\label{fig:pspec}%
Combined proton spectrum from the decay of \sili{}.
Spectra are shown for $\Delta E$-$E$ detector telescopes no. 3 (59 µm $\Delta E$) and 4 (300 µm $\Delta E$); see Table \ref{tab:silicons}.
In the spectra, the initial proton kinetic energies, $E_p$, have been converted to observed proton center-of-mass energies, $E_{p}^{\mathrm{cm}}$.
The incomplete regions (IR; see section \ref{subsubsec:reconstruction}) for the two detector telescopes are indicated; they span roughly from 2.2 to 2.8 MeV and from 5.9 to 7.1 MeV, respectively.
The onset of the incomplete telescope regions explains the artificial structure in those energy regions.
A $\beta$-gated spectrum for telescope no. 3 is also shown; see text.
}
\end{figure*}

A sample of the proton spectra obtained in the present experiment are shown in Fig. \ref{fig:pspec}.
The spectra shown are for $\Delta E$-$E$ detector telescopes no. 3 and 4 (Table \ref{tab:silicons}), where the 300 µm detector $\Delta E$-4 is the detector with superior resolution across the largest energy range, until punch through starts occurring and the acceptance drops across the incomplete telescope region, as explained in section \ref{subsubsec:reconstruction}.
The detector $\Delta E$-3 has comparable resolution below its punch through threshold, but once the reconstructed energy spectrum becomes a combination with the energy observed in the backing $E$-3 detector, the resolution deteriorates a bit.
As was mentioned in section \ref{subsubsec:silicon-cal}, we believe that more statistics in the internal calibration could have yielded more accurate energy calibrations for the backing $E$ detectors, since a more elaborate energy calibration scheme would then be possible.

Fig. \ref{fig:pspec} demonstrates how the low-energy protons emitted in the decay of \sili{} are only cleanly observed by employing thin $\Delta E$ detectors, such as $\Delta E$-3.
In the figure, $\beta$-particles that also go through $E$-3 will give rise to summing, as seen for the two intense IAS  proton peaks at around 4.2 and 5.6 MeV; requiring the $\beta$-particle to be seen in another $E$ detector of the setup (energy deposition above low-energy threshold) removes these sum peaks.

\begin{table*}
\caption{\label{tab:protons}%
$\beta$-delayed protons in the decay of \sili{}.
The observed proton center-of-mass energies, $E_{p}^{\mathrm{cm}}$, are presented for each transition to the $i$'th state in \magn{} and attributed to the same level of excitation energy $E_{\mathrm{ex}}$ in \alum{}.
The deduced average proton center-of-mass energies relative to the $0^+$ ground state in \magn{}, $E_{p,0}^{\mathrm{cm}}$, are listed to the left along with the deduced level widths, $\Gamma$.
The values of $E_{\mathrm{ex}}$ are calculated based on $E_{p,0}^{\mathrm{cm}}$.
The proton branching ratios, $b_i$, relative to the strongest peak at \mbox{$E_{p}^{\mathrm{cm}} =$ 4262 keV} are given along with penetrabilities $P_\ell$ for different angular momentum values $\ell$.
The values of projection, $K$, of total angular momentum along the symmetry axis in rotational states of \alum{} are adapted from literature, while new spin assignments, $J$, deduced in the present work, are also listed.
}
\begin{ruledtabular}
\begin{tabular}{cccclllllcc}
$E_{\mathrm{ex}}$ (keV) & $E_{p,0}^{\mathrm{cm}}$ (keV) & $\Gamma$ (keV) & \magn{} state $i$ & $E_{p}^{\mathrm{cm}}$ (keV) & $b_i$ (\%) & $P_0$ & $P_2$ & $P_4$ & $K$ & $J$\\
\colrule
2672(3) & 401(3) & $^a$ & 0 & 401(3)$^c$ &  $>58^{d}$ & -- & $8.4 \times 10^{-7}$ & $2.7 \times 10^{-10}$ & 1/2 & \\
3856(4) & 1585(4) & $^a$ & 0 &  1585(4) &  2.1(3) & -- & 0.013 & $3.8 \times 10^{-5}$ & 1/2  & \\
4193(4) & 1922(4) & $<30$ & 0 & 1921(4)$^c$ & 24.3(10) & -- & 0.033 & $1.4 \times 10^{-4}$ &  & \\
& & & 2 & 554(9) & 2.6(5) & $8.5 \times 10^{-4}$ & $1.5 \times 10^{-5}$&  &  & \\
4581(4) & 2310(4) & $<30$ & 0 & 2309(4)$^c$ & 14.4(7) & -- & 0.071 & $4.6 \times 10^{-4}$ & & \\
& & & 2 & 942(4)$^c$ & 15(1) & 0.023 & $6.7 \times 10^{-4}$& & & \\
4902(6) & 2631(6) & $^a$ & 0 & -- & & -- & 0.118 & $1.0 \times 10^{-3}$ & 1/2 & (7/2) \\
& & & 2 & 1262(6) & 2.4(3) & 0.092 & $4.0 \times 10^{-3}$& & & \\
5802(4) & 3531(4) & $<30$ & 0 & 3520(15) & 4(3) & -- & 0.312 & $5.7 \times 10^{-3}$ & & 3/2 \\
& & & 2 & 2162(4)$^c$ & 15.8(7) & 0.487 & 0.054 & & & \\
6116(8) & 3845(8) & 60(20) & 0 & 3850(10) & 1.2(6) & -- & 0.397 & $9.1 \times 10^{-3}$ & 3/2 & \\
& & & 2 & 2470(10) & 2.1(8) & 0.651 & 0.094 & & & \\
6654(10) & 4383(10) & 60(20) & 0 & 4385(15) & 1.0(4) & -- & 0.553 & 0.018 & 3/2 & \\
& & & 2 & 3014(10) & 3.8(6) & 0.914 & 0.188 & & & \\
6875(6) & 4604(6) & 30(10) & 0 & -- & & -- & 0.624 & 0.023 & & (7/2) \\
& & & 2 & 3235(6) & 3.1(8) & 1.023 & 0.239 & & & \\
6965(15) & 4694(15) & 80(30) & 0 & -- & & -- & 0.645 & 0.025 &  & 3/2 \\
& & & 2 & 3325(15) & 4.3(10) & 1.053 & 0.254 & & & \\
7115(7) & 4844(7) & 60(20) & 0 & 4847(8) & 9.3(10) & -- & 0.698 & 0.030 &  & \\
& & & 2 & 3472(7)$^c$ & 34.8(10) & 1.129 & 0.296 & & & \\
7244(5) & 4973(5) & 30(10) & 0 & 4974(8) & 2.0(4) & -- & 0.740 & 0.034 & &  \\
& & & 2 & 3604(5) & 6.6(8) & 1.187 & 0.330 & & & \\
& & & $2_2$ & 729(10) & $<0.1$ &  & $5.5 \times 10^{-3}$ & $1.2 \times 10^{-4}$ & &  \\
7425(5) & 5154(5) & $<30$ & 0 & -- & & -- & 0.797 & 0.040 & 3/2 & 7/2 \\
& & & 2 & -- & & 1.264 & 0.378 & & & \\
& & & 4 & 1031(5) & 0.7(1) & 0.036 & $1.2 \times 10^{-3}$ & $1.5 \times 10^{-6}$ & & \\
7721(50) & 5450(50) & $>60$ & 0  & 5450(50) & 0.4(2) & & & &  & \\
7903(5) & 5632(5) & $^a$ & 0 & 5634(6)$^c$ & 22.0(6) & -- & 0.952 & 0.061 & 3/2 & \\
& & & 2 & 4262(5)$^c$ & 100 & 1.460 & 0.518 & & & \\
& & & 4 & 1512(8) & 1.9(6) & -- & 0.010 & $2.6 \times 10^{-5}$ & & \\
& & & $2_2$ & 1391(7) & 3.2(3) & 0.133 & $6.7 \times 10^{-3}$ & & & \\
8192(4) & 5921(4) & $^a$ & 0 & -- & -- & -- & 1.044 & 0.077 & & 7/2 \\
& & & 2 & 4554(10) & 1.6(6) & 1.570 & 0.606 & & & \\
& & & 4 & 1797(4) & 6.6(5) & 0.303 & 0.023 & $8.8 \times 10^{-5}$ & & \\
& & & $2_2$ & 1682(4) & 1.6(3) & 0.249 & 0.017 & & & \\
9008(10) & 6737(10) & 35(15) & 0 &  6750(15) & (0.7) & -- & 1.303 & 0.135 &  &  \\
& & & 2 & 5365(10) & 2.1(4) & 1.858 & 0.867 & & & \\
& & & 4 & 2610(10) & 1.0(3) & 0.719 & 0.115 & $1.0 \times 10^{-3}$ & & \\
9076(10) & 7000(25) & $^b$ & 0 & 7000(25) & $^b$ &  --- & 1.385 & 0.159 & & \\
9271(25) & 6805(10) & 30(15) & 0 & 6805(10) & 0.24(8) & -- & 1.322 & 0.141 & & \\
9965(15) & 7694(15) & $^a$ & 0 & 7693(15) & 0.009(5) & -- & 1.592 & 0.231 & & \\
& & & 2 & 6325(15) & 0.03(1) & 2.158 & 1.174 & & & \\
\end{tabular}
\end{ruledtabular}
\begin{flushleft}
$^a$Width less than 25 keV.\\
$^b$See text for a discussion of the high energy region.\\
$^c$Transition used for internal calibration of silicon detectors.\\
$^d$Lower limit; see text.
\end{flushleft}
\end{table*}

We list in Table \ref{tab:protons} the experimentally determined energies and branching ratios.
For each level in \alum{}, a common level position (quoted as the proton center-of-mass energy relative to the $0^+$ ground state of \magn{}, $E_{p,0}^{\mathrm{cm}}$) is extracted; these are the same values given in the right-most column of Table \ref{tab:peaks}.
The quoted level energies, $E_{\mathrm{ex}}$, of \alum{} in Table \ref{tab:protons} are calculated based on the quoted $E_{p,0}^{\mathrm{cm}}$.
The 401 keV proton line is affected by detection thresholds, as mentioned in section \ref{subsubsec:silicon-cal}, so only a lower limit of its intensity is reported.
In the following, we refer to the observed proton center-of-mass energies, $E_{p}^{\mathrm{cm}}$, when discussing proton peaks, and we refer to excitation energies in \alum{}, $E_{\mathrm{ex}}$, when discussing level energies.

Four levels listed in \cite{Bas25} as being fed in $\beta$-decay are not confirmed in our data:
(1) The proton peak at 3326 keV earlier assumed to go to the ground state is here shown to proceed to the $2^+$ state, so there is no evidence of $\beta$-feeding to the 5597 keV level.
(2) The proton peak which we find at 3845 keV has in earlier experiments \cite{Hat92,Rob93,Tho04} been reported to originate from a level at 6123 keV, 6131 keV or 6170 keV; it must belong to the $3/2^+$ 6121 keV level, so the 6170 keV level is not confirmed here.
(3) A similar discrepancy occurs for the proton peak at 3014 keV going to the $2^+$ state in $^{24}$Mg where \cite{Hat92} agrees with our assignment to the $5/2^+$ 6649 keV level, whereas the position at 6620 keV reported in \cite{Tho04} is not confirmed.
(4) The proton peak at 5369 keV earlier assumed to go to the ground state is here shown to proceed to the $2^+$ state, so there is no evidence of $\beta$-feeding to the 7646 keV level.
Furthermore, the spectral shape around 7144 keV has the shape of a plateau rather than a peak, so we also cannot yet confirm the level at 9415 keV.
The high energy part of the proton spectrum is discussed in more detail in section \ref{subsubsec:high-energy}.

The known level at 6518 keV would have a ground state transition peak at 4247 keV, which is hidden below the main proton peak in the spectrum.
A transition to the first excited state would correspond to a peak at 2878 keV, but no clear signal is seen there either within our statistics.
For these reasons, we cannot set a realistic upper limit for a $\beta$-transition to this level.
The possible transition to the 7971 keV level is discussed below.

Our proton energy resolution along with the $\gamma$-coincidences indicates that two proton peaks are previously unresolved double peaks:
The first is the proton peak at 1500 keV where both components go to excited states in $^{24}$Mg; this peak will be discussed further in section \ref{subsubsec:beta-p-gamma}.
The second is the prominent peak slightly below 3500 keV, the main component is the excited state transition from the 7112 keV level, but a ground state component from the 5804 keV level (as discussed earlier \cite{Ree66,Hat92}) probably contributes in the upper half of the peak.

The ground state transition corresponding to the $2^+$ transition at 3235 keV is not observed, but it could lie beneath a stronger peak.
Although the ground state transition from the known broad 7716 keV level also lies below more intense peaks, we believe to have indications for it, whereas the evidence for the corresponding transition to the first excited state is too weak.

Detailed fits to the proton line shapes have not been carried out, so the level widths, $\Gamma$, in Table \ref{tab:protons} have conservative uncertainties.
Apart from broadening due to the detector resolution, the natural line shape is also broadened due to $\beta$-recoil.
The latter effect will in general make the peaks slightly more round around their maxima and contribute with a spread in position that can range from 4 to 9 keV (standard deviation).
For most peaks, the broadening will be in the range 7--9 keV, but transitions to the $4^+$ and second $2^+$ states in \magn{} can have smaller contributions.

Ref.\ \cite{Hat92} suggested that interference effects, which would manifest differently in the $0^+$ ground state and $2^+$ transitions, could explain why the 7112 keV level seemed to differ in widths in the two instances.
The peak for the excited state transition is, as mentioned above, part of an unresolved double peak and we find that the widths agree when this is taken into account.
However, several of the peaks in the spectrum are rather wide and interference may be visible in the tail regions.
We comment briefly on this here.

For the spin 3/2 levels, the transition of the 5804 keV level to the $2^+$ state (the peak at 2162 keV) is asymmetric; there is a sharp fall-off on the high energy side.
This may indicate interference with the $3/2^+$ 6121 keV level, so we assign the same spin to the 5804 keV level.
In a similar manner, the 6944 keV and 7112 keV levels appear to interfere constructively at energies between them for the transitions to the $2^+$ state; both levels then have spin 3/2.
To get a sharp fall off to higher energies, for the ground state transition peaks, there must also be interference with the broad 7716 keV level.
The interference with the 7971 keV level (if it is fed) will be too small to be observed, but one would expect that the tail upwards from the 7716 keV level will be rather long.

For the spin 5/2 levels there could be constructive interference between the 6649 keV and 7242 keV level, explaining the rapid fall-off above the 4974 keV peak.
The IAS is too narrow and located too far from these levels to display any features.
Also, the spin 7/2 levels are too narrow and too far apart to give rise to visible interference.

\subsubsection{The $\beta$p$\gamma$-decays}\label{subsubsec:beta-p-gamma}

\begin{figure*}
\includegraphics[width=1.00\textwidth]{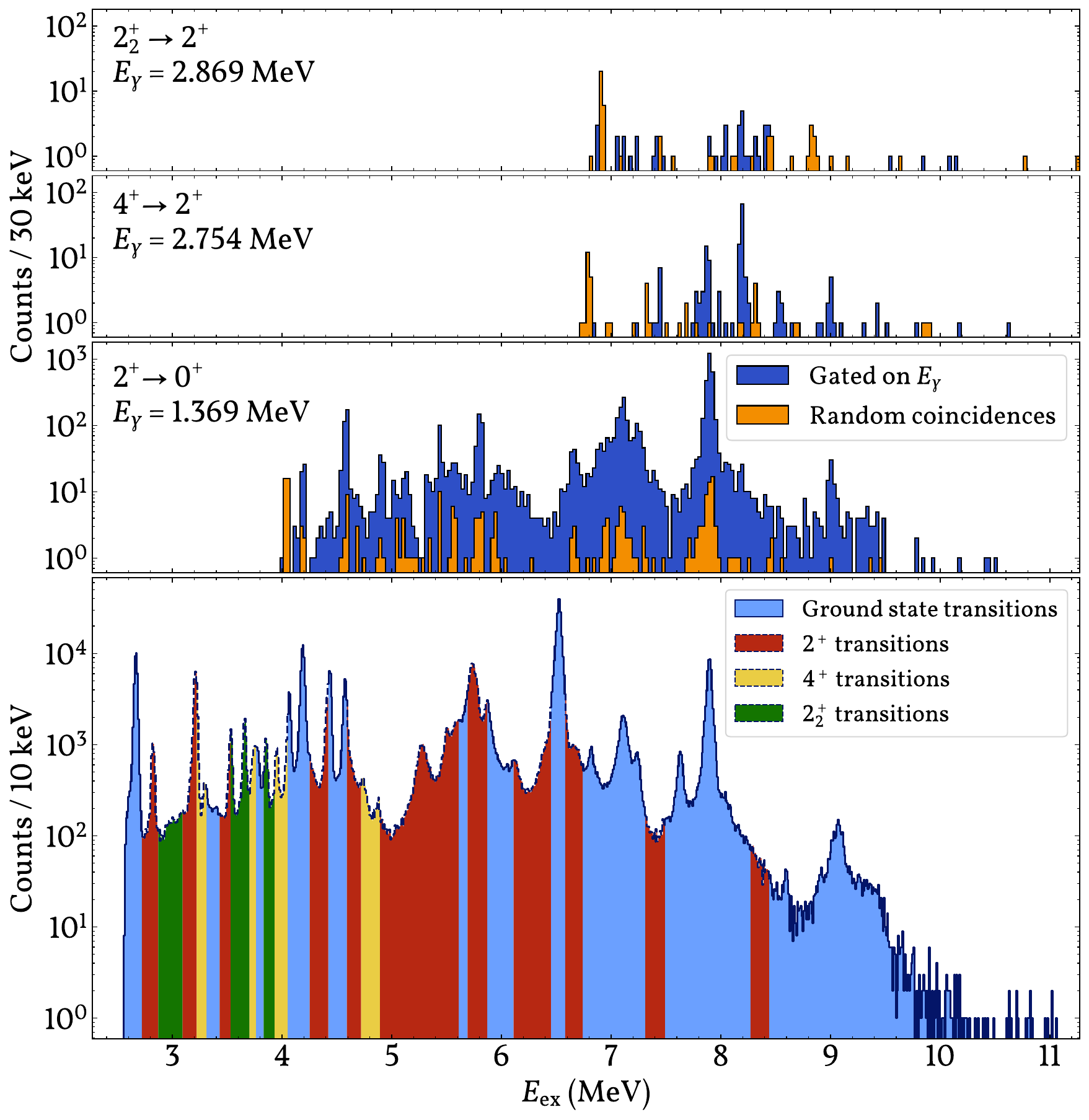}
\caption{\label{fig:pgspec}%
Combined proton spectrum from the decay of \sili{} plotted against the deduced excitation energy, $E_{\mathrm{ex}}$, in \alum{}.
The bottom panel assumes the protons are emitted to the ground state of \magn{} and distinguishes between ground state transitions as well as transitions attributed to excited states in \magn{}.
The spectra in the upper panels are gated on $\gamma$-ray energies, $E_{\gamma}$, corresponding to the indicated transitions in \magn{} (see Fig. \ref{fig:decayscheme}) and are offset by the excitation energies of the initial states in the indicated transitions.
}
\end{figure*}

The range of $\gamma$-ray energies recorded in our setup and the resolution obtained was, as explained in section \ref{subsec:detectors}, limited.
We therefore only recorded coincidences between protons and the $\gamma$-rays at 1369 keV, 2754 keV and 2869 keV; see Figs. \ref{fig:gamma-spectra} and \ref{fig:pgspec}.
The efficiency for getting a proton-$\gamma$ coincidence is around 1.2\% at 1369 keV and 0.8\% at 2.8 MeV.
Fortunately, the earlier detailed \sili{} decay spectroscopy experiment \cite{Sun21} did a very careful analysis of the $\gamma$ emission in the decay and could also perform proton-$\gamma$ coincidences up to a proton energy of around 2 MeV, which is exactly the region where protons going to the $4^+$ and second $2^+$ state in \magn{} are found.

Since the $4^+$ state decays through the first $2^+$ state, we see transitions to this state in coincidence with both $\gamma$ transitions.
Only four proton lines are seen to decay in this manner.
The second $2^+$ state decays 78\% of the time to the ground state and 22\% of the time through the first $2^+$ state.
Hence, we have a reduced probability of seeing transitions to the second $2^+$ state in coincidence with the 1369 keV $\gamma$-ray.
We note that the following two states at higher excitation energy in $^{24}$Mg, the $3^+$ state at 5235 keV and the $4_2^+$ state at 6010 keV, would mainly decay through the $2^+$ level, and that transitions to them could not be distinguished from transitions directly to the first $2^+$ level in our data.
We agree with the coincidences reported in \cite{Sun21}, but have improved precision on the proton energies.

We find, in agreement with \cite{Sun21}, that the broad 724 keV peak (already hinted at in \cite{Ree66}) has $\gamma$ coincidences indicating a transition from the 7242 keV level to the $2^+_2$ state; however, its width, as measured in \cite{Ste24} and here, is more consistent with a width clearly larger than 19(4) keV, so it is possible that more than one peak is present in this region.

The proton peak at 1500 keV is wider than the surrounding peaks and its composition is not immediately obvious.
The upper part fits in energy with being the transition from the IAS to the $4^+$ state, and we do, as \cite{Sun21}, find coincidences with the 1369 keV and 2754 keV $\gamma$-rays, as expected.
However, we also find coincidences with the 2869 keV $\gamma$-ray that could indicate feeding of the second $2^+$ state but do not see other proton lines that could arise from the same level in \alum{}.
We tentatively assign the upper half of the peak as arising from the IAS, but more statistics will be needed to get a clear identification.

There is an indication in Fig. 4 in \cite{Sun21} for a proton-coincident $\gamma$-ray at 3866 keV, the main line from the $3^+$ state.
If this is true, its width corresponds to low-energy protons and its relative intensity is of order 0.4.
This may fit either with proton emission from the (new, see below) level at 9008 keV where three other transitions are seen (the proton peak energy of the transition to the $3^+$ state is then 1502 keV) or it could result from a transition from the IAS to the $3^+$ state, giving a proton peak at 397 keV that lies below the very intense peak at 401 keV.

Table \ref{tab:protons} includes penetrabilities for the transitions calculated for different angular momenta (employing a radius of $1.4\times(24^{1/3}+1^{1/3})$ fm $=$ 5.44 fm).
If the levels in \alum{} have total angular momenta $3/2^+$ or $5/2^+$, protons are emitted with orbital angular momenta $\ell =$ 0 for transitions to the $2^+$ state and $\ell =$ 2 to the $0^+$ and $4^+$ states, whereas levels with angular momenta $7/2^+$ have $\ell =$ 0 to the $4^+$ state, $\ell =$ 2 to the $2^+$ states and $\ell =$ 4 to the $0^+$ ground state.
The patterns in penetrabilities can be compared to the observed branching ratios.
The 7421 keV level is clearly seen in p-$\gamma$ coincidence to the $4^+$ state, both in \cite{Sun21} and in the present experiment.
Our data could be consistent with a weak transition to the $2^+$ state occuring close to a proton peak energy of 3785 keV.
This decay pattern supports the previous assignment of the level spin as $7/2^+$.
In a similar way, $7/2^+$ is the favored spin for the 8189 keV level that mainly decays to the $4^+$ state, and our data would suggest that the levels at 4906 keV and 6879 keV also have spin $7/2^+$.

We did not find many earlier reports on the ratios of proton emission to the different states in $^{24}$Mg.
We are in reasonable agreement on the ratio $b_2/b_0$ with the reaction work in \cite{Pri91} for the levels at 6121 keV, 7242 keV and 7716 keV, but not for the 7112 keV level.

\subsubsection{The high-energy region}\label{subsubsec:high-energy}

\begin{figure}
\includegraphics[width=1.00\columnwidth]{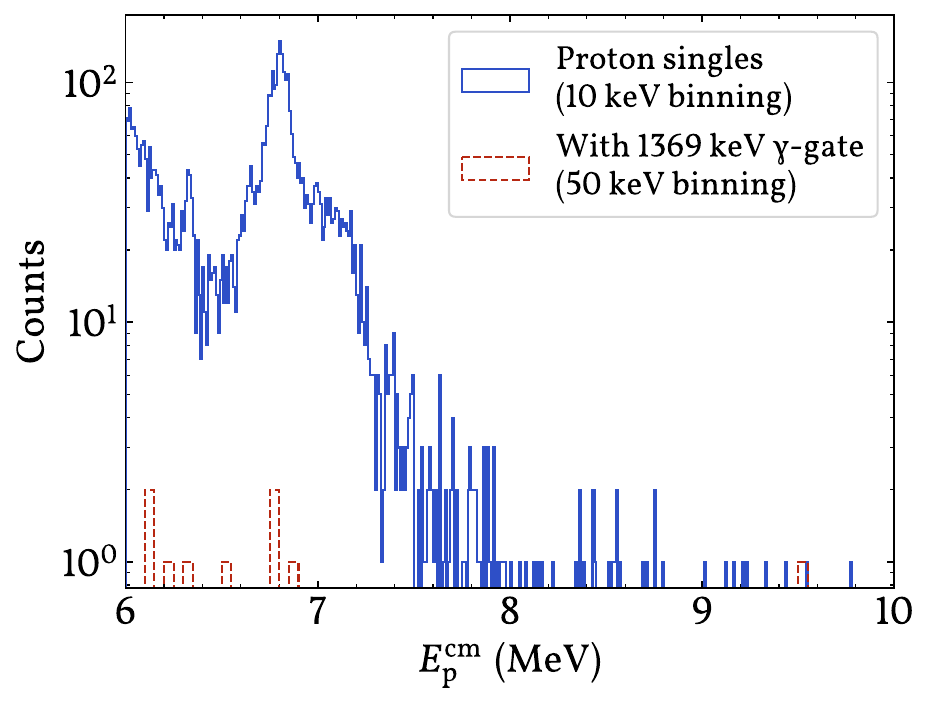}
\caption{\label{fig:pEabove6}%
High-energy part of the combined proton spectrum from the decay of \sili{}.
The initial proton kinetic energies, $E_p$, have been converted to observed proton center-of-mass energies, $E_{p}^{\mathrm{cm}}$.
Events coincident with a 1369 keV $\gamma$-ray are shown in red.
}
\end{figure}

For the high-energy region, taken here to be above the 8189 keV level, there is less information available in the literature.
The most informative earlier $\beta$-decay work for this energy region is in \cite{Zho85} which lists three peaks above the IAS.
We confirm the two lowest of their quoted peaks, but see a more complex structure in this region, as shown in Fig. \ref{fig:pEabove6}.

We find the main peak at high energy to be positioned at a center-of-mass energy of 6805 keV.
However, the intensity on its lower flank must be due to a new peak that we place at 6737 keV, i.e. a level energy of 9008 keV.
Transitions from this level to the $2^+$ and $4^+$ states in \magn{} are recorded as well.
We note that the shape at the lower edge of the ground state and first excited state peaks are quite similar.
Above the 6805 keV peak there is a plateau structure where there could be indications for a peak around 7000 keV, but the higher lying structure is more complex.
We note that there are coincidences corresponding to transitions to the first excited state from the upper part of this structure (but not the 7000 keV peak) and that its energy overlaps with the highest peak of \cite{Zho85}.

There is an indication for a peak around 7370 keV, but more statistics will be needed to make reliable assignments for the very highest energies.
The only tentative assignment we shall propose is that of a peak at 9965 keV.
That would give rise to the clearly visible peak at proton center-of-mass energy 6325 keV.
A single event has a 1369 keV $\gamma$-ray in coincidence, which fits with an indication for a peak at 7693 keV, so we interpret this peak as going to the first excited state; for a peak of this size one coincident event would be expected whereas random coincidences are unlikely.
Other coincidences with a 1369 keV $\gamma$-ray are present in this region, but cannot yet be attributed.

\begin{figure*}
\includegraphics[width=1.00\textwidth]{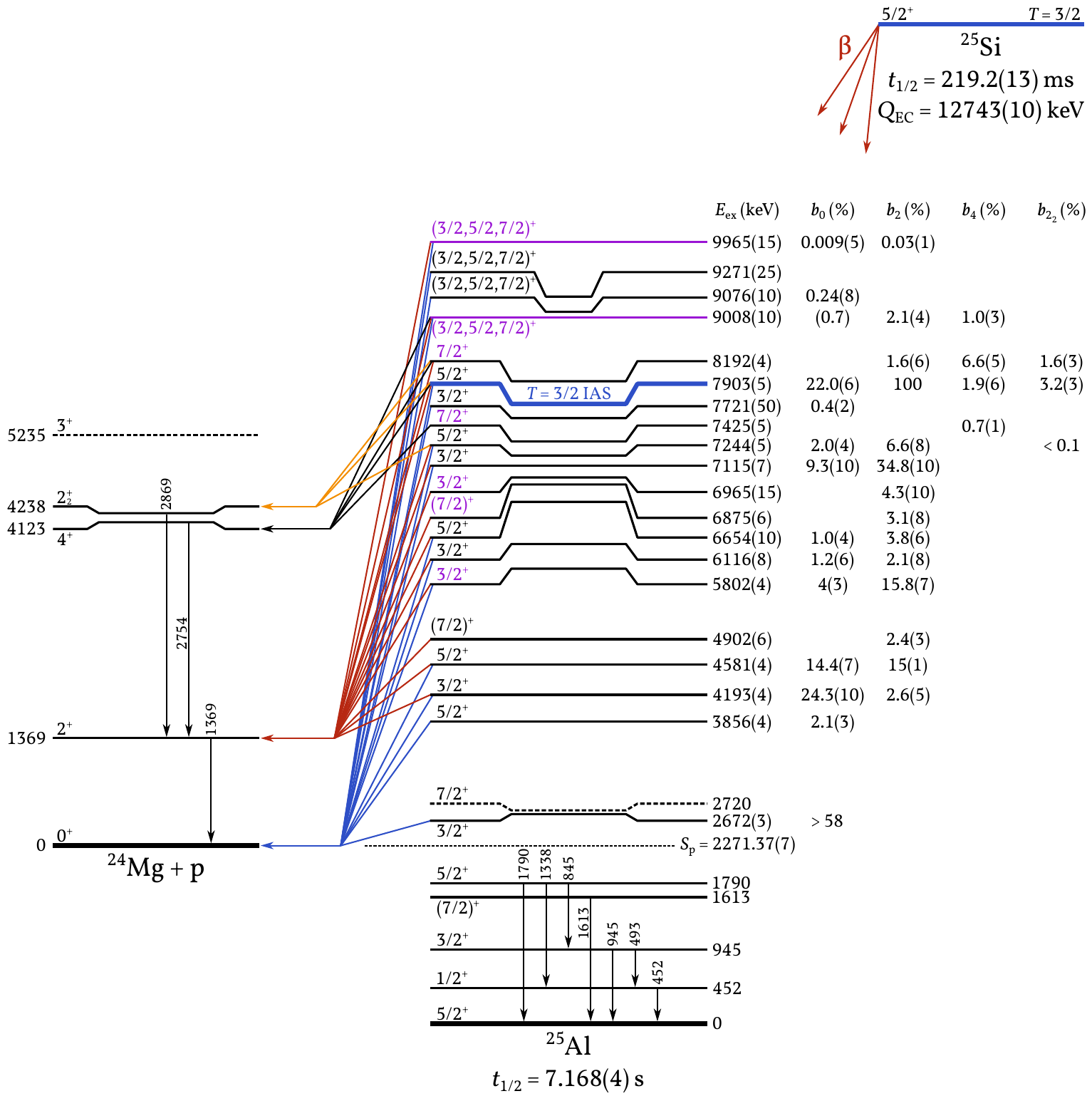}
\caption{\label{fig:decayscheme}%
Decay scheme of \sili{}, summarizing the results presented in Table \ref{tab:protons}.
Excitation energies, $E_{\mathrm{ex}}$, are given in keV along with the branching ratios, $b_i$, relative to the strongest peak from the IAS at $E_{\mathrm{ex}} =$ 7903 keV to the $2^+$ excited state in \magn{} at 1369 keV.
The excitation energies for proton transitions are calculated as described in Table \ref{tab:protons}.
The remaining excitation energies are taken from literature.
New level assignments at high excitation energies in \alum{} are highlighted in purple, as are new spin assignments.
}
\end{figure*}

A decay scheme of \sili{}, summarizing all findings of the present data, is shown in Fig. \ref{fig:decayscheme}.
New level assignments at high excitation energies in \alum{} are highlighted in purple, as are new spin assignments.
Arguments for the new spin assignments are further substantiated in the next section.

\subsection{Comments on the data}\label{subsec:comments}

It is tempting to compare the decay pattern of \sili{} with that of $^{21}$Mg as the nuclei involved mainly differ by the addition of an $\alpha$ particle.
We observe in both cases that the proton emissions from the daughter nuclei go preferentially to the $2^+$ excited state in $^{20}$Ne and $^{24}$Mg, respectively, rather than to the $0^+$ ground states.

The consequences of the proton emission taking place between deformed nuclei were investigated in \cite{Rii24}.
There, general expressions for the intensity of proton emission between deformed nuclei were derived and applied to the case of the $K=1/2$ nuclide $^{21}$Mg.
The final expressions for the intensity of proton emission includes sums over several Clebsch-Gordan coefficients that, apart from a $K$-selection rule (introduced in section \ref{subsec:structure}), implies a preference for certain decay paths.
The $K$-selection rule, for the relevant case of \sili{}, dictates that $s$-wave proton emission to the $K=0$ ground state band of \magn{} can only happen from states in \alum{} with $K = 1/2$.
The $K$-selection rule is not strictly followed, but hinders transitions when it is violated.

The Clebsch-Gordan coefficients in the expressions for the intensity of proton emission introduce preferences in state transitions based on initial and final state total angular momenta, $J$, as well as initial and final $K$ quantum numbers:
For levels with $J=7/2$, proton emission to the $0^+$ ground state is suppressed; if the levels furthermore have $K=1/2$, emission to the $2^+$ state is also suppressed.
For levels with $J=3/2$, there is a similar suppression of emission to the ground state for $K=1/2$; only slightly less so for $K=3/2$.
Finally, for levels with $J=5/2$, there is little $J$-preference for $K=1/2$, a moderate suppression of transitions to the lowest $2^+$ state for $K=3/2$, and a similar suppression of transitions to the $4^+$ state for $K=5/2$.
On top of this overall ``geometric'' suppression of emission to the $0^+$ state, there may also be structure effects that favour the $2^+$ state due to the (radial) overlap matrix elements, since wavefunctions may be dominated by $d$-wave components.

The just outlined decay preferences fit nicely with the suggested assignments of $J=7/2$ levels in Table \ref{tab:protons}, but suggests that the two highest such levels, at 7421 keV and 8189 keV, have $K=1/2$.
The decay pattern of the IAS is in reasonable agreement with a $J=5/2$, $K=3/2$ assignment, but points to a structural enhancement of decays to the lowest $2^+$ state; a similar conclusion holds for the 6649 keV level.
The decay pattern of the new 9008 keV level should be established better before spin assignments are suggested.

There has so far been no report of $\beta$-decay transitions to the isospin $T=3/2$, $J=3/2$ level at 7971 keV.
It is located 68 keV above the IAS and we do not see any indications of proton emission from it.
It is expected to decay mainly via $\gamma$ emission \cite{Bas25}, but was not observed in the $\gamma$ spectrum in \cite{Sun21}.
A conservative upper limit for its intensity is 1\% of the IAS intensity.

\section{The $\beta$-strength distribution}

The intensity distribution can be converted to a $\beta$-strength distribution, as is done e.g.\ in \cite{Jen24}.
The $\beta$-strength at a given excitation energy $E_{\mathrm{ex}}$ with corresponding branching ratio $BR$ is calculated as

\begin{equation}
B_{\beta}(E_{\mathrm{ex}})=\frac{{BR}\,\mathcal{T}_{1/2}}{f\,t_{1/2}}
\label{eq:beta}
\end{equation}
where $\mathcal{T}_{1/2} = 6144.48\pm3.70$ s \cite{Har20}, $t_{1/2} = 219.2\pm1.3$ ms is the half-life of \sili{} \cite{Bas25} and $f$ is the phase space factor calculated here with the BetaShape program \cite{Mou19}, assuming an uncertainty on $f$ of 1\%.
For states below the proton separation energy in \alum{} we employ branching ratios from \cite{Sun21}.
For the 401 keV line just above the proton separation energy, we employ a relative branching ratio of 74\%, as given in \cite{Rob93}.

\begin{figure*}
\includegraphics[width=1.00\textwidth]{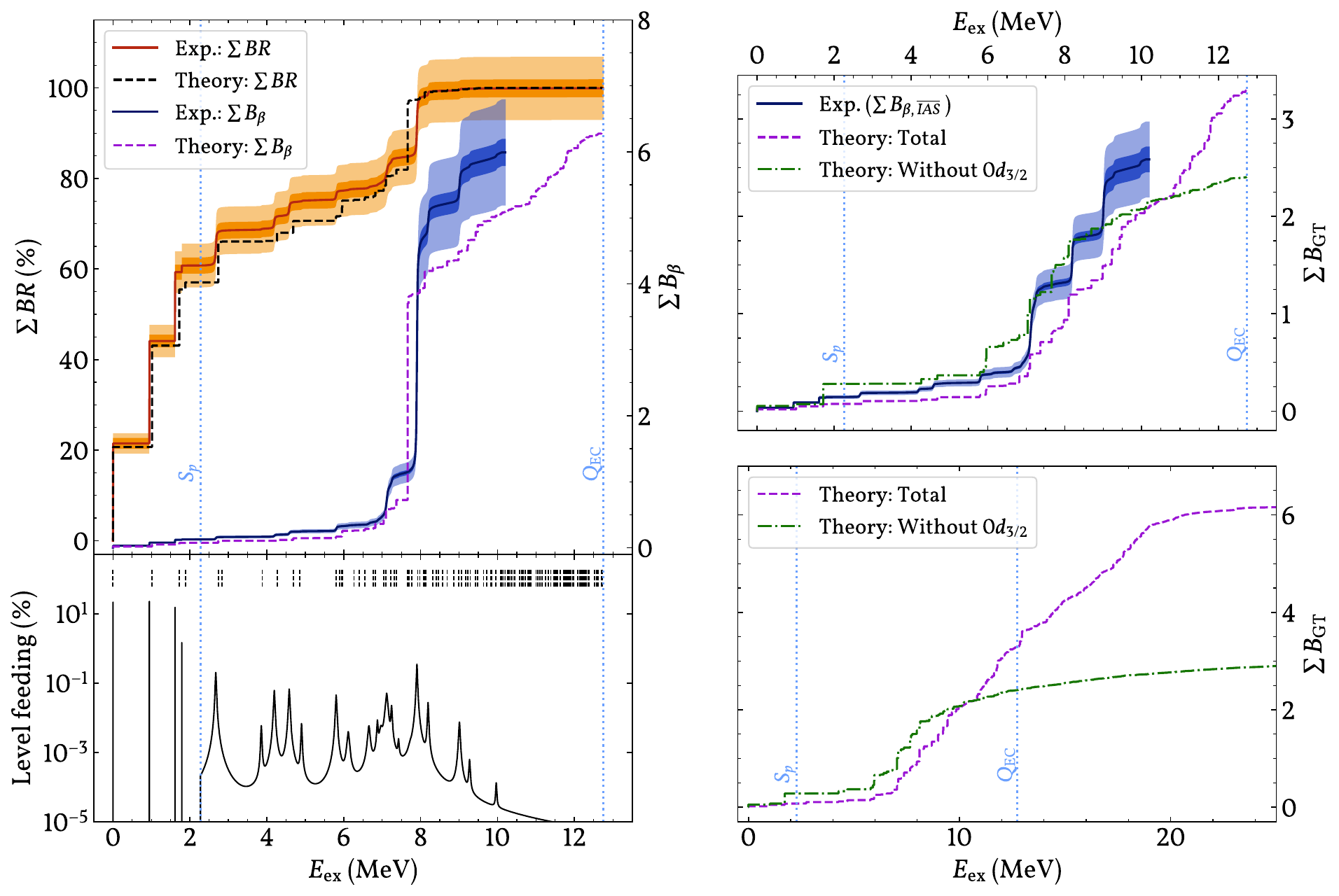}
\caption{\label{fig:Bbeta}%
Cumulative $\beta$-strength and cumulative branching ratios as a function of excitation energy $E_{\mathrm{ex}}$ in \alum{}.
The proton separation energy, $S_p$, and electron capture Q-value, $Q_{\mathrm{EC}}$, are indicated.
\textbf{Top left:} Experimental and theoretical distributions of total cumulative $\beta$-strength, $\sum{B_\beta}$, and branching ratios, $\sum{BR}$.
The blue band and the dark orange band illustrate statistical uncertainties on $\sum{B_\beta}$ and $\sum{BR}$, respectively.
The light blue band and light orange band illustrate the estimated systematic uncertainties of respectively 5\% and 10\%.
\textbf{Bottom left:} Theoretical level feedings (dashed vertical lines) are drawn above experimental level feedings (solid line).
The experimental level feedings above $S_p$ are modeled as isolated Breit-Wigner resonances.
\textbf{Top right:} Cumulative GT strength, $\sum{B_{\mathrm{GT}}}$, from theory is compared to experimental cumulative $\beta$-strength excluding the IAS contribution, $\sum{B_{\beta,\overline{IAS}}}\, 	\simeq\sum{B_{\mathrm{GT}}}$.
Theoretical $\sum{B_{\mathrm{GT}}}$ distributions are plotted with and without $0d_{3/2}$ components; see text.
\textbf{Bottom right:} Theoretical $\sum{B_{\mathrm{GT}}}$ distributions are plotted up to $E_{\mathrm{ex}} =$ 25 MeV in \alum{}.
}
\end{figure*}

The cumulative $\beta$-strength and cumulative branching ratios in the decay of \sili{} are presented in Fig. \ref{fig:Bbeta}.
Experimental results are compared to theoretical predictions based on ``universal'' $sd$-shell model calculations \cite{Mag20}.
The dark orange band surrounding the red curve and the blue bands surrounding the dark blue curves illustrate statistical uncertainties on the cumulative branching ratios and cumulative $\beta$-strength, respectively.
The light orange band and light blue bands illustrate the estimated systematic uncertainties of respectively 5\% and 10\% on the cumulative branching ratios and cumulative $\beta$-strength.
The statistical uncertainties are estimated based on Monte Carlo simulations, taking into account the uncertainties on the individual branching ratios as well as on $\mathcal{T}_{1/2}$, $t_{1/2}$ and $f$ in Eq. (\ref{eq:beta}).

The theoretical calculations were carried out in the $sd$-shell model space with the USDC Hamiltonian \cite{Mag20} using the shell-model code NuShellx@MSU \cite{Bro14}.
The Gamow-Teller strength, $B_{\mathrm{GT}}$, values were reduced by a factor of $R = $ 0.60.
This represents the typical average quenching factor from Table III of \cite{Ric08} ($Q = q_{\mathrm{GT}}^{2}$).
The $B_{\mathrm{GT}}$ from the present data require $R_{\mathrm{exp}} =$ 0.75, which is within the variation observed between experiment and theory for other $B_{\mathrm{GT}}$ values (Fig. 8 of \cite{Ric08}).
The total summed $B_{\mathrm{GT}}$ shown in the bottom right of Fig. \ref{fig:Bbeta} extends up to 25 MeV.
The summed strength is $\Sigma B_{\mathrm{GT^+}} =$ 6.2 compared to the Ikeda sum rule of $(\Sigma  B_{\mathrm{GT^+}} - \Sigma  B_{\mathrm{GT^-}}) R = 3(Z-N) R =$ 5.4, with $R\,\Sigma B_{\mathrm{GT^-}} =$ 0.8 and $R =$ 0.6.

The largest partition in the \sili{} ground-state wavefunction is 36\% for $[0d_{5/2}]^{9}$ with other components spread out over about 300 other partitions.
The Gamow-Teller strength obtained by leaving out the $0d_{3/2}$ components of the one-body transition density is shown by the green dash-dotted line in Fig. \ref{fig:Bbeta}.
This shows that the low-lying Gamow-Teller strength observed below 6 MeV is a result of a strong cancellation between $0d_{5/2}$ and $0d_{3/2}$ contributions.

The reduction factor of 0.60 for the Gamow-Teller strength from \cite{Ric08} comes mainly from the decay of nuclei near stability.
A connection has been suggested between the reduction in the Gamow-Teller strength and the reduction of spectroscopic factors obtained for example from (e,e'p) experiments relative to those obtained from shell-model calculations \cite{Cau95}.
In \cite{Tos14} it has been shown that the experimental spectroscopic quenching depends on the distance from stability through the factor for proton removal as measured by the difference in proton and neutron separation energies, $\Delta S = S_p-S_n$.
For \sili{}, $\Delta S = -12$ MeV leading to a reduction factor of about 0.8(1), referring to Fig. 1 in \cite{Tos14}.
Perhaps the smaller reduction of 0.75 (compared to the ``standard'' quenching of 0.60) which we observe in the Gamow-Teller strength is connected to the observations made in \cite{Cau95,Tos14}.
More analysis and experiments on the $\beta$-decay of proton-rich nuclei are needed to confirm this connection.

We note that the total experimental $\beta$-strength to the IAS is 3.7(4), possibly larger than the total Fermi strength of 3.
The theoretical Gamow-Teller strength to the IAS is 0.093.
A sizable Gamow-Teller component in the IAS could be probed through the $\beta$-recoil broadening, which can be assessed either by a detailed line shape analysis or through measurement of the kinematic shift of protons versus the proton-$\beta$-angle.
An initial analysis of the present data \cite{Ede24} indicates that more statistics will be needed to get an interesting level of precision on such studies.

\section{Conclusion}

The production yields and selectivity of modern in-flight facilities allow for detailed decay studies of proton-rich nuclei, as exemplified with the previous $\gamma$-ray study of $^{25}$Si \cite{Sun21}.
However, a reliable and sensitive study of particle emission in the decays needs very thin sources, as demonstrated in the present paper.
Furthermore, we have demonstrated that the combination of ARIS, the ACGS and compact high-granularity detector arrays enables precision stopped-beam charged-particle spectroscopy at FRIB.
The achieved resolution and sensitivity establish a new powerful platform for future studies of increasingly exotic proton-rich nuclei.
In this context, \sili{} is an excellent test case as well as a solid reference nuclide for calibration purposes with its broad range of intense proton and $\gamma$ lines that are known to high precision and accuracy.

Although \sili{} in the present experimental campaign mainly serves as a calibration source, the excellent resolution in $\beta$-delayed proton spectra along with the $\gamma$-proton coincidences (supplemented with the data from \cite{Sun21}) has allowed for a critical revision of the decay scheme.
Several reassignments and new assignments has led to an experimental $\beta$-strength distribution extending up to 10 MeV excitation energy, approaching the electron capture Q-value.
The experimental $\beta$-strength distribution agrees with state-of-the-art configuration-interaction calculations in the $sd$-shell model space, although the need for a smaller reduction in the theoretical Gamow-Teller strength suggests that more studies of the $\beta$-decays of proton-rich nuclei are needed to understand the evolution of the quenched $\beta$-strength towards the proton drip line.

\begin{acknowledgments}

The authors acknowledge the support from the Independent Research Fund Denmark, Projects No. 2032-00066B, 4283-00172B, and 9040-00076B, from the Swedish Research Council, Project No. 2022-04248, from the U.S. National Science Foundation under Grants No. PHY-1913554, PHY-2110365, PHY-2209429, and PHY-2514797, from the Spanish MICIU/AEI/10.13039/501100011033, and from FEDER, EU, under Project No. PID2022-140162NB-I00.

This material is based upon work supported by the U.S. Department of Energy, Office of Science, Office of Nuclear Physics and used resources of the Facility for Rare Isotope Beams (FRIB) Operations, which is a DOE Office of Science User Facility under Award Number DE-SC0023633.

\end{acknowledgments}

\bibliography{main}

@ARTICLE{Sun25,
   author       = "L. J. Sun and J. Dopfer and A. Adams and C. Wrede and A. Banerjee and B. A. Brown and J. Chen and E. A. M. Jensen and R. Mahajan and T. Rauscher and C. Sumithrarachchi and L. E. Weghorn and D. Weisshaar and T. Wheeler",
   title        = "{Extension of the particle x-ray coincidence technique: The lifetimes and branching ratios apparatus}",
   journal      = "Phys.\ Rev.\ C",
   volume       = "111",
   pages        = "055806",
   year         = "2025",
   url          = "https://doi.org/10.1103/PhysRevC.111.055806"
}

@ARTICLE{Bas22,
   author       = "M. S. Basunia and A. Chakraborty",
   title        = "{Nuclear Data Sheets for A=24}",
   journal      = "Nucl.\ Data\ Sheets",
   volume       = "186",
   pages        = "3",
   year         = "2022",
   url          = "https://doi.org/10.1016/j.nds.2022.11.002"
}

@ARTICLE{Bas25,
   author       = "M. S. Basunia and A. Chakraborty",
   title        = "{Nuclear Structure and Decay Data for A=25 Isobar}",
   journal      = "Nucl.\ Data\ Sheets",
   volume       = "205",
   pages        = "1",
   year         = "2025",
   url          = "https://doi.org/10.1016/j.nds.2025.08.001"
}

@ARTICLE{Por23,
   author       = "M. Portillo and B. M. Sherrill and Y. Choi and M. Cortesi and K. Fukushima and M. Hausmann and E. Kwan and S. Lidia and P. N. Ostroumov and R. Ringle and M. K. Smith and M. Steiner and O. B. Tarasov and A. C. C. Villari and T. Zhang",
   title        = "{Commissioning of the Advanced Rare Isotope Separator ARIS at FRIB}",
   journal      = "Nucl.\ Instrum.\ Methods\ Phys.\ Res.\ B",
   volume       = "540",
   pages        = "151",
   year         = "2023",
   url          = "https://doi.org/10.1016/j.nimb.2023.04.025"
}

@ARTICLE{Lun20,
   author       = "K. R. Lund and G. Bollen and D. Lawton and D. J. Morrissey and J. Ottarson and R. Ringle and S. Schwarz and C. S. Sumithrarachchi and A. C. C. Villari and J. Yurkon",
   title        = "Online tests of the {A}dvanced {C}ryogenic {G}as {S}topper at {NSCL}",
   journal      = "Nucl.\ Instrum.\ Methods\ Phys.\ Res.\ B",
   volume       = "463",
   pages        = "378",
   year         = "2020",
   url          = "https://doi.org/10.1016/j.nimb.2019.04.053"
}

@ARTICLE{Jen23,
   author       = "E. A. M. Jensen and K. Riisager and H. O. U. Fynbo",
   title        = "{Extracting clean low-energy spectra from silicon strip detector telescopes around punch through energies}",
   journal      = "Nucl.\ Instrum.\ Methods\ Phys.\ Res.\ A",
   volume       = "1055",
   pages        = "168531",
   year         = "2023",
   url          = "https://doi.org/10.1016/j.nima.2023.168531"
}

@ARTICLE{Zie10,
   author       = "J. F. Ziegler and M. D. Ziegler and J. P. Biersack",
   title        = "{SRIM} – The stopping and range of ions in matter (2010)",
   journal      = "Nucl.\ Instrum.\ Methods\ Phys.\ Res.\ B",
   volume       = "268",
   pages        = "1818",
   year         = "2010",
   url          = "https://doi.org/10.1016/j.nimb.2010.02.091"
}

@ARTICLE{Hua21,
   author       = "W. J. Huang and M. Wang and F. G. Kondev and G. Audi and S. Naimi",
   title        = "{The AME2020 atomic mass evaluation (I). Evaluation of input data, and adjustment procedures}",
   journal      = "Chin.\ Phys.\ C",
   volume       = "45",
   pages        = "030002",
   year         = "2021",
   url          = "https://doi.org/10.1088/1674-1137/abddb0"
}

@ARTICLE{Wan21,
   author       = "M. Wang and W. J. Huang and F. G. Kondev and G. Audi and S. Naimi",
   title        = "{The AME2020 atomic mass evaluation (II). Tables, graphs and references}",
   journal      = "Chin.\ Phys.\ C",
   volume       = "45",
   pages        = "030003",
   year         = "2021",
   url          = "https://doi.org/10.1088/1674-1137/abddaf"
}

@ARTICLE{Mar67,
   author       = "M. A. Mariscotti",
   title        = "{A method for automatic identification of peaks in the presence of background and its application to spectrum analysis}",
   journal      = "Nucl.\ Instrum.\ Methods",
   volume       = "50",
   pages        = "309",
   year         = "1967",
   url          = "https://doi.org/10.1016/0029-554X(67)90058-4"
}

@ARTICLE{Len86,
   author       = "W. N. Lennard and H. Geissel and K. B. Winterbon and D. Phillips and T. K. Alexander and J. S. Forster",
   title        = "{Nonlinear response of Si detectors for low-Z ions}",
   journal      = "Nucl.\ Instrum.\ Methods\ Phys.\ Res.\ A",
   volume       = "248",
   pages        = "454",
   year         = "1986",
   url          = "https://doi.org/10.1016/0168-9002(86)91033-8"
}

@ARTICLE{Hat92,
   author       = "S. Hatori and H. Miyatake and S. Morinobu and K. Katori and M. Fujiwara and I. Katayama and N. Ikeda and T. Fukuda and T. Shinozuka and K. Ogawa",
   title        = "{Gamow-Teller strength in $\beta$-decay of $^{25}$Si}",
   journal      = "Nucl.\ Phys.\ A",
   volume       = "549",
   pages        = "327",
   year         = "1992",
   url          = "https://doi.org/10.1016/0375-9474(92)90083-V"
}

@ARTICLE{Rob93,
   author       = "J. D. Robertson and D. M. Moltz and T. F. Lang and J. E. Reiff and J. Cerny",
   title        = "{Beta-delayed proton decay of $^{25}$Si}",
   journal      = "Phys.\ Rev.\ C",
   volume       = "47",
   pages        = "1455",
   year         = "1993",
   url          = "https://doi.org/10.1103/PhysRevC.47.1455"
}

@ARTICLE{Tho04,
   author       = "J.-C. Thomas and L. Achouri and J. Äystö and R. Béraud and B. Blank and G. Canchel and S. Czajkowski and P. Dendooven and A. Ensallem and J. Giovinazzo and N. Guillet and J. Honkanen and A. Jokinen and A. Laird and M. Lewitowicz and C. Longour and F. {de Oliveira Santos} and K. Peräjärvi and M. Stanoiu",
   title        = "{Beta-decay properties of $^{25}$Si and $^{26}$P}",
   journal      = "Eur.\ Phys.\ J.\ A",
   volume       = "21",
   pages        = "419",
   year         = "2004",
   url          = "https://doi.org/10.1140/epja/i2003-10218-8"
}

@ARTICLE{Sun21,
   author       = "L. J. Sun and M. Friedman and T. Budner and D. {Pérez-Loureiro} and E. Pollacco and C. Wrede and B. A. Brown and M. Cortesi and C. Fry and B. E. Glassman and J. Heideman and M. Janasik and A. Kruskie and A. Magilligan and M. Roosa and J. Stomps and J. Surbrook and P. Tiwari",
   title        = "{$^{25}$Si $\beta^+$-decay spectroscopy}",
   journal      = "Phys.\ Rev.\ C",
   volume       = "103",
   pages        = "014322",
   year         = "2021",
   url          = "https://doi.org/10.1103/PhysRevC.103.014322"
}

@ARTICLE{Ste24,
   author       = "I. C. Stefanescu and L. Trache and A. Saastamoinen and B. T. Roeder and A. E. Spiridon and A. I. Stefanescu and E. Pollacco and G. Lotay and R. E. Tribble",
   title        = "{$\beta$-delayed proton decay of $^{27}$P and $^{25}$Si: Implications for the $^{26m}$Al(p,$\gamma$)$^{27}$Si reaction rate}",
   journal      = "Phys.\ Rev.\ C",
   volume       = "110",
   pages        = "015804",
   year         = "2024",
   url          = "https://doi.org/10.1103/PhysRevC.110.015804"
}

@ARTICLE{Rii24,
   author       = "K. Riisager and E. A. M. Jensen and A. S. Jensen",
   title        = "{Beta-delayed particle emission and collective rotations}",
   journal      = "Eur.\ Phys.\ J.\ A",
   volume       = "61",
   pages        = "87",
   year         = "2025",
   url          = "https://doi.org/10.1140/epja/s10050-025-01550-4"
}

@ARTICLE{Jen24,
   author       = "E. A. M. Jensen and S. T. Nielsen and A. Andreyev and M. J. G. Borge and J. Cederkäll and L. M. Fraile and H. O. U. Fynbo and L. J. Harkness-Brennan and B. Jonson and D. S. Judson and O. S. Kirsebom and R. Lică and M. V. Lund and M. Madurga and N. Marginean and C. Mihai and R. D. Page and A. Perea and K. Riisager and O. Tengblad",
   title        = "{Detailed study of the decay of $^{21}$Mg}",
   journal      = "Eur.\ Phys.\ J.\ A",
   volume       = "60",
   pages        = "153",
   year         = "2024",
   url          = "https://doi.org/10.1140/epja/s10050-024-01376-6"
}

@ARTICLE{Fuj04,
   author       = "Y. Fujita and I. Hamamoto and H. Fujita and Y. Shimbara and T. Adachi and G. P. A. Berg and K. Fujita and K. Hatanaka and J. Kamiya and K. Nakanishi and Y. Sakemi and Y. Shimizu and M. Uchida and T. Wakasa and and M. Yosoi",
   title        = "{Evidence for the Existence of the [2 0 2]3/2 Deformed Band in Mirror Nuclei $^{25}$Mg and $^{25}$Al}",
   journal      = "Phys.\ Rev.\ Lett.",
   volume       = "92",
   pages        = "062502",
   year         = "2004",
   url          = "https://doi.org/10.1103/PhysRevLett.92.062502"
}

@ARTICLE{Ree66,
   author       = "P. L. Reeder and A. M. Poskanzer and R. A. Esterlund and R. McPherson",
   title        = "{Beta-Delayed Protons from Si$^{25}$}",
   journal      = "Phys.\ Rev.",
   volume       = "147",
   pages        = "781",
   year         = "1966",
   url          = "https://doi.org/10.1103/PhysRev.147.781"
}

@ARTICLE{Lon18,
   author       = "B. Longfellow and A. Gade and B. A. Brown and W. A. Richter and D. Bazin and P. C. Bender and M. Bowry and B. Elman and E. Lunderberg and D. Weisshaar and S. J. Williams",
   title        = "{Measurement of key resonances for the $^{24}$Al(p,$\gamma$)$^{25}$Si reaction rate using in-beam $\gamma$-ray spectroscopy}",
   journal      = "Phys.\ Rev.\ C",
   volume       = "97",
   pages        = "054307",
   year         = "2018",
   url          = "https://doi.org/10.1103/PhysRevC.97.054307"
}

@ARTICLE{Ell68,
   author       = "J. P. Elliott and C. E. Wilsdon",
   title        = "{Collective motion in the nuclear shell model IV. Odd-mass nuclei in the $sd$ shell}",
   journal      = "Proc.\ Roy.\ Soc.\ A",
   volume       = "302",
   pages        = "509",
   year         = "1968",
   url          = "https://doi.org/10.1098/rspa.1968.0033"
}

@ARTICLE{Jan73,
   author       = "J. Jänecke",
   title        = "{The reaction $^{26}$Mg(d, $\tau$)$^{25}$Na and the structure of $^{25}$Na}",
   journal      = "Nucl.\ Phys.\ A",
   volume       = "204",
   pages        = "497",
   year         = "1973",
   url          = "https://doi.org/10.1016/0375-9474(73)90390-4"
}

@ARTICLE{Col75,
   author       = "B. J. Cole and A. Watt and R. R. Whitehead",
   title        = "{Shell-model calculations in the sd shell III. The structure of mass 25 nuclei}",
   journal      = "J.\ Phys.\ G",
   volume       = "1",
   pages        = "17",
   year         = "1975",
   url          = "https://doi.org/10.1088/0305-4616/1/1/005"
}

@ARTICLE{Zho85,
   author       = "Z. Y. Zhou and E. C. Schloemer and M. D. Cable and M. Ahmed and J. E. Reiff and J. Cerny",
   title        = "{Additional beta-delayed protons from the $T_z=-3/2$ nuclei $^{21}$Mg, $^{25}$Si, $^{29}$S and $^{41}$Ti}",
   journal      = "Phys.\ Rev.\ C",
   volume       = "31",
   pages        = "1941",
   year         = "1985",
   url          = "https://doi.org/10.1103/PhysRevC.31.1941"
}

@ARTICLE{Pri91,
   author       = "R. M. Prior and S. E. Darden and K. R. Nyga and H. {Paetz Gen. Schieck}",
   title        = "{States of $^{25}$Al between 6 and 8 MeV}",
   journal      = "Nucl.\ Phys.\ A",
   volume       = "533",
   pages        = "411",
   year         = "1991",
   url          = "https://doi.org/10.1016/0375-9474(91)90525-B"
}

@book{Boh75,
  title={Nuclear Structure, Volume II: Nuclear Deformations},
  author={A. Bohr and B. R. Mottelson},
  year={1975},
  publisher={W. A. Benjamin, Inc.},
  address={Reading, Massachusetts},
  isbn={0805310169}
}

@PHDTHESIS{Jens24,
   author  = "E. A. M. Jensen",
   title   = "{Decays At The Edge. Probing Light Proton-Rich Nuclei at State-Of-The-Art Rare Ion Beam Facilities}",
   type    = "{PhD} Dissertation",
   address = "Aarhus University",
   year    = "2024",
   url     = "https://repository.cern/records/8sbt0-hht05"
}

@PHDTHESIS{Ede24,
   author  = "J. M. Eder",
   title   = "{Beta Decay 25Si. An FRIB Experimental Approach}",
   type    = "Master's thesis",
   address = "Aarhus University",
   year    = "2024",
   url     = "https://wiki.kern.phys.au.dk/justusME.pdf"
}

@PHDTHESIS{Ped25,
   author  = "P. H. Pedersen",
   title   = "{Beta Decay of the Proton-Rich Nucleus $^{22}$Al}",
   type    = "Master's thesis",
   address = "Aarhus University",
   year    = "2025",
   url     = "https://wiki.kern.phys.au.dk/PhilipSpeciale.pdf"
}

@ARTICLE{Fuk23,
   author       = "K. Fukushima and M. Cortesi and M. Hausmann and E. Kwan and P. N. Ostroumov and M. Portillo and B. M. Sherrill and M. Smith and M. Steiner and T. Zhang",
   title        = "{Simulation studies for beam commissioning at FRIB Advanced Rare Isotope Separator}",
   journal      = "Nucl.\ Instrum.\ Methods\ Phys.\ Res.\ B",
   volume       = "541",
   pages        = "53",
   year         = "2023",
   url          = "https://doi.org/10.1016/j.nimb.2023.04.038"
}

@ARTICLE{Rin21,
   author       = "R. Ringle and G. Bollen and K. Lund and C. Nicoloff and S. Schwarz and C. S. Sumithrarachchi and A. C. C. Villari",
   title        = "Particle-in-cell techniques for the study of space charge effects in the {A}dvanced {C}ryogenic {G}as {S}topper",
   journal      = "Nucl.\ Instrum.\ Methods\ Phys.\ Res.\ B",
   volume       = "496",
   pages        = "61",
   year         = "2021",
   url          = "https://doi.org/10.1016/j.nimb.2021.03.020"
}

@article{Mag20,
  title = {{New isospin-breaking ``USD'' Hamiltonians for the $sd$ shell}},
  author = {A. Magilligan and B. A. Brown},
  journal = {Phys.\ Rev.\ C},
  volume = {101},
  pages = {064312},
  year = {2020},
  url = {https://doi.org/10.1103/PhysRevC.101.064312}
}

@article{Ten04,
  title = {{Novel thin window design for a large-area silicon strip detector}},
  author = {O. Tengblad and U. C. Bergmann and L. M. Fraile and H. O. U. Fynbo and S. Walsh},
  journal = {Nucl.\ Instrum.\ Methods\ Phys.\ Res.\ A},
  volume = {525},
  pages = {458},
  year = {2004},
  url = {https://doi.org/10.1016/j.nima.2004.01.082}
}

@article{Vin21,
  title = {{Calibration and response function of a compact silicon-detector set-up for charged-particle spectroscopy using GEANT4}},
  author = {S. Viñals and E. Nácher and O. Tengblad and M. J. G. Borge and J. A. Briz and A. Gad and M. Munch and A. Perea},
  journal = {Eur.\ Phys.\ J.\ A},
  volume = {57},
  pages = {49},
  year = {2021},
  url = {https://doi.org/10.1140/epja/s10050-021-00371-5}
}

@article{Jen26,
  title = {{Measurement of the Ground State Spin and Parity of $^{22}\mathrm{Al}$ Disfavors Halo Formation}},
  author = "E. A. M. Jensen and J. S. Nielsen and B. S. O. Johansson and A. Adams and J. Dopfer and C. S. Sumithrarachchi and L. J. Sun and L. E. Weghorn and T. Wheeler and C. Wrede and M. J. G. Borge and O. Tengblad and M. Madurga and B. Jonson and K. Riisager and H. O. U. Fynbo",
  journal = {Phys.\ Rev.\ Lett.},
  volume = {136},
  pages = {202503},
  year = {2026},
  url = {https://doi.org/10.1103/3lpm-sy41}
}

@article{Bor13,
  title = {{Beta-delayed particle emission}},
  author = "M. J. G. Borge",
  journal = {Phys.\ Scr.\ T},
  volume = {152},
  pages = {014013},
  year = {2013},
  url = {https://dx.doi.org/10.1088/0031-8949/2013/T152/014013}
}

@article{Per16,
  title = {{$\beta$-delayed $\gamma$ decay of $^{26}\mathrm{P}$: Possible evidence of a proton halo}},
  author = {D. P\'erez-Loureiro and C. Wrede and M. B. Bennett and S. N. Liddick and A. Bowe and B. A. Brown and A. A. Chen and K. A. Chipps and N. Cooper and D. Irvine and E. McNeice and F. Montes and F. Naqvi and R. Ortez and S. D. Pain and J. Pereira and C. J. Prokop and J. Quaglia and S. J. Quinn and J. Sakstrup and M. Santia and S. B. Schwartz and S. Shanab and A. Simon and A. Spyrou and E. Thiagalingam},
  journal = {Phys.\ Rev.\ C},
  volume = {93},
  pages = {064320},
  year = {2016},
  url = {https://doi.org/10.1103/PhysRevC.93.064320}
}

@ARTICLE{Liu18,
   author       = {Q. Liu and Y. Ye and Z. Li and C. Lin and H. Jia and Y. Ge and Q. Li and J. Lou and X. Yang and B. Yang and J. Feng and H. Zang and Z. Chen and Y. Liu and W. Liu and S. Chen and H. Yu and J. Li and Y. Zhang and F. Yang and L. Yang and N. Ma and L. J. Sun and D. Wang},
   title        = "{Investigation of the thickness non-uniformity of the very thin silicon-strip detectors}",
   journal      = "Nucl.\ Instrum.\ Methods\ Phys.\ Res.\ A",
   volume       = "897",
   pages        = "100",
   year         = "2018",
   url          = "https://doi.org/10.1016/j.nima.2018.04.041"
}

@article{Pur17,
author = {S. Purushothaman and S. {Ayet San Andrés} and J. Bergmann and T. Dickel and J. Ebert and H. Geissel and C. Hornung and W. R. Plaß and C. Rappold and C. Scheidenberger and Y. K. Tanaka and M. I. Yavor},
title = {Hyper-EMG: A new probability distribution function composed of Exponentially Modified Gaussian distributions to analyze asymmetric peak shapes in high-resolution time-of-flight mass spectrometry},
journal = {Int.\ J.\ Mass\ Spectrom.},
volume = {421},
pages = {245},
year = {2017},
url = {https://doi.org/10.1016/j.ijms.2017.07.014},
}

@article{Har20,
  title = {Superallowed ${0}^{+}\rightarrow{0}^{+}$ nuclear $\beta$ decays: 2020 critical survey, with implications for ${V}_{\mathit{ud}}$ and CKM unitarity},
  author = {J. C. Hardy and I. S. Towner},
  journal = {Phys. Rev. C},
  volume = {102},
  pages = {045501},
  year = {2020},
  url = {https://doi.org/10.1103/PhysRevC.102.045501}
}

@article{Mou19,
  title = {Towards high-precision calculation of electron capture decays},
  journal = {Appl.\ Radiat.\ Isot.},
  volume = {154},
  pages = {108884},
  year = {2019},
  url = {https://doi.org/10.1016/j.apradiso.2019.108884},
  author = {X. Mougeot}
}

@article{Bro14,
  title = {The Shell-Model Code NuShellX@MSU},
  journal = {Nucl.\ Data\ Sheets},
  volume = {120},
  pages = {115},
  year = {2014},
  url = {https://doi.org/10.1016/j.nds.2014.07.022},
  author = {B. A. Brown and W. D. M. Rae}
}

@article{Ric08,
  title = {$\mathit{sd}$-shell observables for the USDA and USDB Hamiltonians},
  author = {W. A. Richter and S. Mkhize and B. A. Brown},
  journal = {Phys.\ Rev.\ C},
  volume = {78},
  pages = {064302},
  year = {2008},
  url = {https://doi.org/10.1103/PhysRevC.78.064302}
}

@article{Cau95,
  title = {Missing and Quenched Gamow-Teller Strength},
  author = {E. Caurier and A. Poves and A. P. Zuker},
  journal = {Phys.\ Rev.\ Lett.},
  volume = {74},
  pages = {1517},
  year = {1995},
  url = {https://doi.org/10.1103/PhysRevLett.74.1517}
}

@article{Tos14,
  title = {Systematics of intermediate-energy single-nucleon removal cross sections},
  author = {J. A. Tostevin and A. Gade},
  journal = {Phys.\ Rev.\ C},
  volume = {90},
  pages = {057602},
  year = {2014},
  url = {https://doi.org/10.1103/PhysRevC.90.057602}
}

@article{Bro90,
  title = {Isospin-forbidden \ensuremath{\beta}-delayed proton emission},
  author = {B. A. Brown},
  journal = {Phys.\ Rev.\ Lett.},
  volume = {65},
  pages = {2753},
  year = {1990},
  url = {https://doi.org/10.1103/PhysRevLett.65.2753}
}

\appendix

\section{Further technical details on setup}\label{app:detector-details}

This appendix provides a few more details on the employed germanium and silicon detectors described in section \ref{subsec:detectors}.

The germanium crystals in the two detectors are of diameter 84.8 mm and 79.8 mm, respectively, and of length 65.2 mm and 80.0 mm, respectively.
Both crystals were housed behind a 0.60 mm carbon window, and the distance from the crystal faces to the center of the thin catcher foil was $88 \pm 10$ mm, where the uncertainties are conservative estimates in terms of alignment of the germanium detectors with respect to the center of the detector holder (parallel to the beam axis) and the distance between the vacuum chamber and the faces of the germanium detectors (orthogonal to the beam axis).

The dead layers on the junction sides of the $\Delta E$ detectors consist of 0.1 µm $p^+$ doping and 0.2 µm aluminum arranged in a thin grid such that only 3-4\% of the detector surface is covered with the aluminum readout contacts.
On the ohmic side, the dead layers consist of 0.4 µm $n^+$ doping and 0.2 µm aluminum readout contacts which entirely cover each strip.
The dead layers on both the junction and the ohmic sides of the $E$ detectors consist of 0.4 µm doping and 0.2 µm aluminum readout contacts.

The distance from the center of the detector holder to the centers of each of the $n^+$-side surfaces of the 4 W1 ($\Delta E$) detectors in the plane is 41.8 mm.
The $n^+$-side surfaces of the MSX25 ($E$) detectors in the plane are situated 6.0 mm behind the $n^+$-side surfaces of their corresponding W1 detector.
In the top and bottom faces of the detector holder, the distances from the center of the detector holder to the $n^+$-side surfaces of the W1 detectors are 41.0 mm.
In the bottom face, the distance from the $n^+$-side surface of the W1 detector to the $n^+$-side surface of the MSX25 detector is 5.3 mm.

\section{$\Delta E$ detector thickness estimation}\label{app:thickness-estimation}

\begin{figure}
\includegraphics[width=0.90\columnwidth]{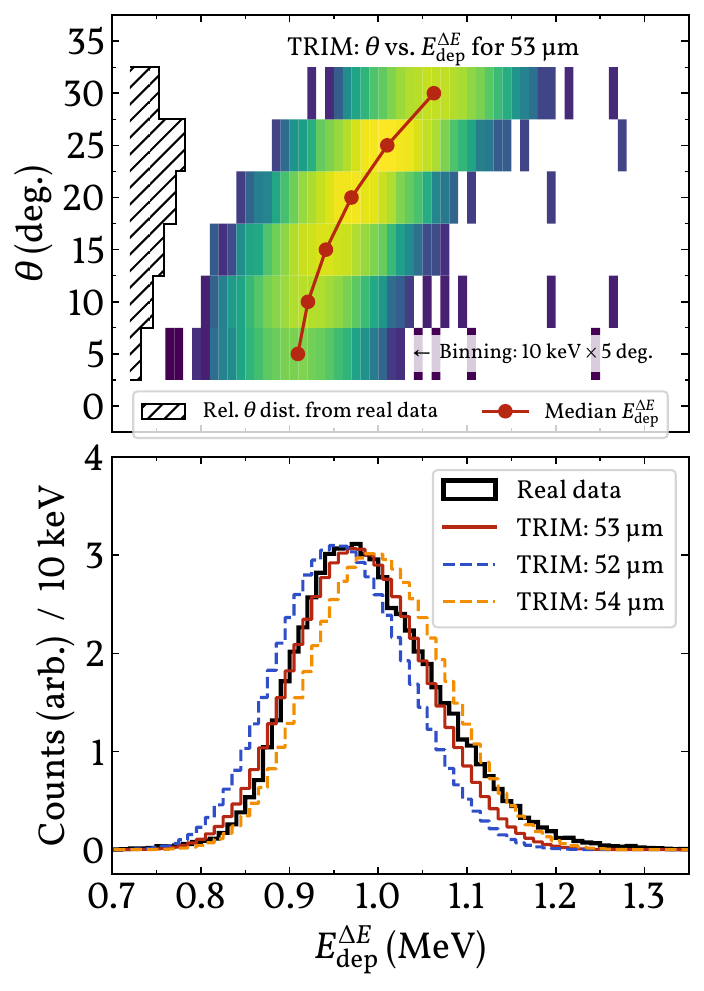}
\caption{\label{fig:thickness-estimation}%
Thickness estimation of $\Delta E$-1 (Table \ref{tab:silicons}) from comparison of real and simulated deposited energy, $E_{\mathrm{dep}}^{\Delta E}$, in \mbox{$\Delta E$-1}.
A well-known reference proton peak from the decay of \sili{} ($p_7$ in Table \ref{tab:25si-cal}) punches through $\Delta E$-1 and is fully stopped in the backing detector, $E$-1, depositing a bit less than 1 MeV in $\Delta E$-1 on average.
\textbf{Top:} Simulated relative distributions of $E_{\mathrm{dep}}^{\Delta E}$ in 53 µm of silicon within various ranges of angles of incidence with respect to detector surface normal, $\theta$.
The effective thickness of the silicon scales as $1/\cos{\theta}$.
The simulated distributions are normalized to the relative experimental distribution of $\theta$, illustrated by the dashed histogram, within the reference proton peak.
\textbf{Bottom:} Projection of relative $\theta$ vs. $E_{\mathrm{dep}}^{\Delta E}$ distributions for $\Delta E$ active thicknesses 52, 53 (top panel) and 54 µm onto the $E_{\mathrm{dep}}^{\Delta E}$ axis.
The distributions have been folded with Gaussians with standard deviation 40 keV and are compared to the real distribution of $E_{\mathrm{dep}}^{\Delta E}$ in order to estimate the thickness of $\Delta E$-1.
}
\end{figure}

As mentioned in section \ref{subsubsec:silicon-cal}, accurate knowledge of the thickness of the $\Delta E$ detectors is necessary when carrying out energy calibrations of the $E$ detectors with reference protons which punch through the $\Delta E$ detectors and are subsequently stopped in the corresponding $E$ detectors.
With accurate calibrations of energy depositions, $E_{\mathrm{dep}}$, in both $\Delta E$ and $E$ detectors, it is possible to accurately reconstruct the initial proton kinetic energies, $E_p$, from both types of detectors over the full range of $E_p$.
In order to achieve accurate calibrations of energy depositions in the $E$ detectors, $E_{\mathrm{dep}}^{E}$, accurate calibrations of energy depositions in the $\Delta E$ detectors, $E_{\mathrm{dep}}^{\Delta E}$, as well as accurate estimates of the $\Delta E$ detector thicknesses are needed; the latter also relying on the former.

Fig. \ref{fig:thickness-estimation} shows an example of how the thicknesses of the thin $\Delta E$ detectors employed in the experiment were estimated to micrometer precision:
The reference proton $p_7$ in Table \ref{tab:25si-cal} of initial proton kinetic energy $E_p =$ 4091(2) keV punches through all $\Delta E$ detectors with thicknesses less than 70 µm, depositing a fraction of its energy, before being fully stopped in the backing $E$ detectors.
The profile of the energy deposition of punch through protons in a given $\Delta E$ detector, $E_{\mathrm{dep}}^{\Delta E}$, is indicative of the thickness of the $\Delta E$ detector.
With a fairly large range of angles of incidence with respect to detector surface normal, $\theta$, the overall $E_{\mathrm{dep}}^{\Delta E}$ profile is a convolution of $\theta$-dependent energy deposition profiles, $E_{\mathrm{dep}}^{\Delta E}(\theta)$, since the effective thickness of the $\Delta E$ detector scales as $1/\cos{\theta}$; i.e. for larger $\theta$, the reference proton will deposit more energy on average.
Neglecting (for now) the relevant detector response, the convolution of $\theta$-dependent $E_{\mathrm{dep}}^{\Delta E}(\theta)$ profiles, weighted by the relative intensities of $\theta$ within the energy window of the punch through proton, will be the actual, overall $E_{\mathrm{dep}}^{\Delta E}$ profile.

The Transport of Ions in Matter (TRIM) software \cite{Zie10} was used to simulate energy deposition profiles, $E_{\mathrm{dep}}$, of the reference proton $p_7$ from Table \ref{tab:25si-cal} in different thicknesses of silicon (i.e. the active thickness of the thin $\Delta E$ detectors) at different angles of incidence with respect to surface normal, $\theta$, taking into account the relevant energy losses in preceding inactive media.
An example of such simulations is shown in the top panel of Fig. \ref{fig:thickness-estimation} for 53 µm silicon at various $\theta$.
The energy deposition profiles were weighted by the distribution of $\theta$ within the reference proton peak $p_7$ from real data; an example relative distribution is shown to the left in the top panel of Fig. \ref{fig:thickness-estimation}.

In the bottom panel of Fig. \ref{fig:thickness-estimation}, the resulting weighted distributions of $E_{\mathrm{dep}}^{\Delta E}$ for simulated thicknesses 52, 53 and 54 µm are drawn along with the corresponding $E_{\mathrm{dep}}^{\Delta E}$ profile from real data.
The simulated $E_{\mathrm{dep}}^{\Delta E}$ profiles in the bottom panel of Fig. \ref{fig:thickness-estimation} are folded with Gaussians with standard deviation 40 keV, in order to mimic the dominant part of the detector response of the W1-type detectors with aluminum grid readout contacts \cite{Vin21}.
If the simulated $E_{\mathrm{dep}}^{\Delta E}$ profiles are not folded with Gaussians in this manner, the tails of the distributions drop off faster than observed in the real data while the peak positions of the distributions shift by a few keV.

The profile from the simulated thickness of 53 µm is found to agree the most with the profile from the real data, and the thickness of, in this case, $\Delta E$-1 is estimated to be 53 µm.
The same method was employed to estimate the thicknesses of $\Delta E$-2, $\Delta E$-3 and $\Delta E$-6; see Table \ref{tab:silicons}.
For all detectors, the estimated thicknesses are within 10\% of the thicknesses given in the detectors' specification sheets.

It should be noted that the finite size of the beam spot and the slight variation in implantation spot parallel to the beam axis, as mentioned in relation to Fig. \ref{fig:shadow}, are not taken into account in the simulations.
We also note that the thickness estimation procedure employed here assumes that the $\Delta E$ active layer thicknesses are uniform, i.e.~the non-uniformity of thin silicon detectors investigated e.g. in \cite{Liu18} is not taken into account with this method.
Rather, average active layer thicknesses of the $\Delta E$ detectors are estimated with the method described here.

\end{document}